\documentclass[12pt]{article}
\usepackage{latexsym}
\usepackage{amsmath,,calrsfs}
\usepackage{amsfonts}
\usepackage{amssymb}
\usepackage{amscd}
\usepackage{bbm}
\usepackage{fancybox}
\usepackage{cite}
\usepackage{amsmath,amsfonts,amsbsy}
\usepackage{pstricks,pst-node}
\usepackage[small,bf,hang]{caption2}
\usepackage{graphicx}
\usepackage{epsfig}
\usepackage{psfrag}
\usepackage{comment}

\usepackage{float}

\psset{unit=1.3cm,linewidth=.5pt,radius=.2}  

\usepackage{multirow}                     
\usepackage{float}                          
\usepackage{lscape}                         
\usepackage{bm}

\newcommand{\beq}{\begin{equation}}
\newcommand{\eeq}{\end{equation}}
\newcommand{\bi}{\begin{itemize}}
	\newcommand{\ei}{\end{itemize}}
\newcommand{\bt}{\begin{tabular}}
	\newcommand{\et}{\end{tabular}}
\newcommand{\bc}{\begin{center}}
	\newcommand{\ec}{\end{center}}

\newcommand{\be}{\begin{equation}}
\newcommand{\ee}{\end{equation}}
\newcommand{\bea}{\begin{eqnarray}}
\newcommand{\eea}{\end{eqnarray}}
\newcommand{\ba}{\begin{array}}
	\newcommand{\ea}{\end{array}}

\def\bbox{{\,\lower0.9pt\vbox{\hrule \hbox{\vrule height 0.2 cm
				\hskip 0.2 cm \vrule height 0.2 cm}\hrule}\,}}
\newcommand{\dsl}{\pa \kern-0.5em /}

\makeatletter \@addtoreset{equation}{section} \makeatother

\def\slashchar#1{\setbox0=\hbox{$#1$}           
	\dimen0=\wd0                                 
	\setbox1=\hbox{/} \dimen1=\wd1               
	\ifdim\dimen0>\dimen1                        
	\rlap{\hbox to \dimen0{\hfil/\hfil}}      
	#1                                        
	\else                                        
	\rlap{\hbox to \dimen1{\hfil$#1$\hfil}}   
	/                                         
	\fi}

\begin{document}
	\begin{titlepage}
		\begin{center}
			\vskip 1.5cm
			
			{\large \bf  Exotic Massive 3D Gravity}
			
			\vskip 2cm
			
			{\bf Mehmet Ozkan${}^{1}$, Yi Pang${}^2$,  and
				Paul K.~Townsend${}^3$} \\
			
			\vskip 15pt
			
			{\em $^1$ \hskip -.1truecm
				\em Department of Physics, Istanbul Technical University,\\
				Maslak 34469 Istanbul, Turkey \vskip 5pt }
			
			{email: {\tt ozkanmehm@itu.edu.tr}} \\
			
			\vskip .4truecm
			
			{\em $^2$ \hskip -.1truecm
				\em Mathematical Institute, University of Oxford, \\
				Woodstock Road, Oxford OX2 6GG, U.K. \vskip 5pt }
			
			{email: {\tt Yi.Pang@maths.ox.ac.uk}} \\
			
			\vskip .4truecm
			
			{\em $^3$ \hskip -.1truecm
				\em  Department of Applied Mathematics and Theoretical Physics,\\ Centre for Mathematical Sciences, University of Cambridge,\\
				Wilberforce Road, Cambridge, CB3 0WA, U.K.\vskip 5pt }
			
			{email: {\tt P.K.Townsend@damtp.cam.ac.uk}} \\
			
			\vskip .4truecm
			
		\end{center}
		
		\vskip 0.5cm
		
		\begin{center}
			{\bf ABSTRACT}\\[3ex]
		\end{center}

		The linearized equations of ``New Massive Gravity'' propagate a parity doublet of massive spin-2 modes
		in 3D Minkowski spacetime, but a different non-linear extension is made possible by ``third-way'' consistency.
		There is  a ``Chern-Simons-like'' action,  as for other 3D massive gravity models, but the new 
		theory is ``exotic'':  its action is parity odd.  This ``Exotic Massive Gravity'' is the next-to-simplest case in an infinite sequence
		of third-way consistent 3D gravity theories,  the simplest being the ``Minimal Massive Gravity'' alternative to ``Topologically Massive Gravity''.

	\end{titlepage}
	
	\newpage
	\setcounter{page}{1}
	\tableofcontents
	
	
	\section{Introduction}
	
	The graviton, if it exists \cite{DYSON:2013jra},  is massless, as far as we know \cite{Goldhaber:2008xy}. The main theoretical argument for a strictly massless graviton
	was, for many years, the difficulty  of finding a consistent  interacting field theory that becomes equivalent to the
	Fierz-Pauli  (FP) field theory for a massive spin-2 particle in the linearized limit \cite{Boulware:1973my}.  This theoretical difficulty was overcome  in recent 
	years \cite{deRham:2010ik,deRham:2010kj}, as reviewed in  \cite{Hinterbichler:2011tt}, although other problems remain \cite{Deser:2014fta} and the resulting  ``massive gravity'' models  may
	be ruled out by observational evidence \cite{Bellazzini:2017fep}.  However, the theoretical problem of how to consistently extend the FP theory to include interactions 
	was first solved   in  $2+1$ dimensions (which we abbreviate to 3D)   via the ``New Massive Gravity'' (NMG) model \cite{Bergshoeff:2009hq,Bergshoeff:2009aq,Bergshoeff:2010ad}, and the 
	related parity-violating ``General Massive Gravity'' (GMG) that has a limit to the earlier ``Topologically Massive Gravity'' (TMG) \cite{Deser:1981wh}.  Although such 
	3D massive gravity models have no direct implications for ``real world'' gravity, massive or otherwise, they may have applications to condensed matter physics \cite{Bergshoeff:2017vjg}.

	Omitting a possible cosmological constant term, the NMG field equation  for  the  metric of a three-dimensional (3D) spacetime takes the form\footnote{We use a ``mostly-plus'' metric signature 
	convention, as in \cite{Bergshoeff:2009aq} but  in contrast to the ``mostly-minus'' convention of  \cite{Bergshoeff:2009hq}, which accounts for some sign differences with that work.}
	\begin{equation}\label{NMGeq}
	G_{\mu\nu} - \frac{1}{2m^2} K_{\mu\nu}=0\, ,
	\end{equation}
	where $G_{\mu\nu}$ is the 3D Einstein tensor and $K_{\mu\nu}$ is the tensor obtained by variation with respect to the metric $g_{\mu\nu}$ of a multiple of the integral of the scalar
	$G^{\mu\nu}S_{\mu\nu}$, where $S_{\mu\nu}$ is the 3D Schouten tensor.  This  equation admits a Minkowski
	vacuum, and linearization about it yields an equation for the metric perturbation tensor that  is fourth-order in derivatives. Nevertheless, it is only second-order in time derivatives, and
	is  equivalent to the second-order  FP equation  for a spin-2 particle of mass $m$. Perhaps the simplest way to see this is to observe that the differential subsidiary condition 
	implied by the FP equation can be solved, in 3D, in terms of another symmetric tensor field; when expressed in terms of this new tensor field, the FP equation is precisely
	the linearization of  (\ref{NMGeq})  \cite{Bergshoeff:2009tb}.
	
	Of course, there is an infinite number of  tensors that could be added to the NMG equation without changing the linearized equation in a  Minkowski vacuum. 
	These could arise  from terms in the action that involve higher powers  of the 3D Ricci tensor, but  the full field equation will then typically involve terms that are higher than second-order
	in time derivatives, which will imply the propagation of additional degrees of freedom in non-Minkowski backgrounds, some of which will  be negative energy ``ghosts''.
	
	A simple way to find those 3D massive gravity theories that are guaranteed not to propagate additional unphysical modes is to start from
	a ``Chern-Simons-like'' formulation \cite{Hohm:2012vh,Bergshoeff:2014bia,Merbis:2014vja}. As the name suggests, Chern-Simons-like theories include the 
	dreibein and dual spin-connection one-forms used to construct
	Chern-Simons (CS) actions for 3D gravity \cite{Achucarro:1987vz,MSC,Witten:1988hc,Horne:1988jf}, but they also include additional  auxiliary one-form fields.  An ``$N$-flavour''  
	CS-like action will have  $N$ Lorentz-vector
	one-form fields, of which $(N-2)$ are auxiliary, and the dimension per space point of the physical phase space is (assuming invertibility of the dreibein) $2(N-2)$.
	This is zero for $N=2$ because these are the topological CS cases. The $N=3$ case includes TMG and the $N=4$ case includes NMG; in particular,
	there is a $4$-flavour parity-preserving CS-like action whose field equations reduce to (\ref{NMGeq})  after elimination of  the two
	one-form auxiliary fields and the spin connection.

	Alternatives to both TMG and NMG may be explored in this  framework by considering more general $N=3$ and $N=4$ CS-like models. 
	For $N=3$ one finds in this way the ``Minimal Massive Gravity'' (MMG) theory \cite{Bergshoeff:2014pca,Arvanitakis:2014xna}, which resolves certain 
	difficulties of TMG with an anti-de Sitter (AdS) vacuum.
	The $N=4$ case includes ``Zwei-Dreibein Gravity'' (ZDG) \cite{Bergshoeff:2013xma}; this resolves similar difficulties of NMG  but elimination of the auxiliary
	one-form fields now requires an iterative  procedure \cite{Bergshoeff:2014eca}; this yields an  infinite series of terms in the equation analogous  to (\ref{NMGeq}), with convergence not
	guaranteed for all possible metrics. One aim of this paper is to present a systematic analysis of those  $N=4$ CS-like models for which the auxiliary fields
	may be {\it finitely} eliminated; i.e. without the need to assume the validity of an iterative procedure that generates an infinite series.  With the further restriction to a
	parity-even action, we find {\it only} NMG.
	
	However, a parity-preserving field equation may have a parity-odd action!  The simplest example occurs for 3D CS gravity,  for which there is both a parity even action 
	(equivalent to the 3D Einstein-Hilbert action with a cosmological term) and an `exotic''  parity odd action, which was more recently discussed in \cite{Townsend:2013ela}.
	But that was a topological CS gravity theory. Now, by considering $N=4$ CS-like theories with a parity-odd action, we are led to a similarly ``exotic'' {\it massive} 3D gravity theory
	with (parity-preserving) field equation
	\begin{equation}\label{EMGeq}
	\Lambda g_{\mu\nu} + G_{\mu\nu} - \frac{1}{m^2}H_{\mu\nu} + \frac{1}{m^4} L_{\mu\nu} =0\, ,
	\end{equation}
	where the symmetric traceless $H$-tensor and the symmetric $L$-tensor can be expressed in terms of the Cotton tensor $C$ as follows:
	\begin{equation}\label{HandL}
	H_{\mu\nu}= \epsilon_\mu{}^{\rho\sigma} \nabla_\rho C_{\nu\sigma} \, , \qquad L_{\mu\nu} = \frac12 \epsilon_\mu{}^{\rho\sigma}\epsilon_\nu{}^{\lambda\tau} C_{\rho\lambda}C_{\sigma\tau}\, .
	\end{equation}
	The symbol  $\nabla$ indicates the covariant derivative defined in terms of the standard metric connection, and the alternating tensor $\epsilon$ is defined in 
	terms of the invariant alternating tensor density
	$\varepsilon$ by 
	\begin{equation}
	\sqrt{-\det g}\, \epsilon^{\mu\nu\rho} = \varepsilon^{\mu\nu\rho}\, . 
	\end{equation}
        We should mention here that although it makes sense to set
	$\Lambda=0$ in (\ref{EMGeq}),  this cannot be done in the CS-like action; in this respect,  there is a similarity to the exotic CS action for 3D gravity.  For convenience, we shall refer
	to the new massive 3D gravity theory with field equation (\ref{EMGeq}) as  ``Exotic Massive Gravity'' (EMG).
	
	On dimensional grounds, one might expect the $H$ and $L$ tensors appearing in  (\ref{EMGeq})  to result from variation of some curvature-squared and curvature-cubic terms, 
	respectively, but this is not the case. The reason is simple: neither tensor satisfies the Bianchi-type identity  that is satisfied by any tensor found by variation of an action with 
	respect to the metric. Instead, one finds that
	\begin{equation}\label{non-B}
	\nabla^\mu H_{\mu\nu} \equiv - \epsilon_\nu{}^{\rho\sigma} C_\rho{}^\lambda G_{\sigma\lambda} \, , 
	\qquad \nabla^\mu L_{\mu\nu} \equiv -\epsilon_\nu{}^{\rho\sigma} C_\rho{}^\lambda    H_{\sigma\lambda}\, .
	\end{equation}
	This shows that there  is no action  {\it for the metric alone}  whose variation yields the equation (\ref{EMGeq}). In this respect, EMG is similar to MMG
	and the reason is the same: the consistency of the equation (\ref{EMGeq}) is of ``third way'' type, in the terminology of \cite{Bergshoeff:2015zga}.
	
	To check consistency  with the Bianchi identity satisfied by the Einstein tensor in  (\ref{EMGeq}) we use  (\ref{non-B}) to deduce that
	\begin{equation}\label{consist}
	0= \epsilon_\nu{}^{\rho\sigma} C_\rho{}^\lambda\left(G_{\sigma\lambda} - \frac{1}{m^2} H_{\sigma\lambda} \right) \, .
	\end{equation}
	The fact that the right hand side (RHS)  is not identically  zero calls into question the consistency of  (\ref{EMGeq}) because it appears to imply
	a constraint on curvature that would be incompatible with the propagation of modes by the linearized equation,   but we may now use (\ref{EMGeq})  in (\ref{consist}) to deduce that
	\begin{equation}
	RHS = - \Lambda \epsilon_\nu{}^{\rho\sigma} C_{\rho\sigma} - \frac{1}{m^4} \epsilon_\nu{}^{\rho\sigma} C_\rho{}^\lambda L_{\sigma\lambda} \equiv 0\, ,
	\end{equation}
	and hence that (\ref{consist}) does {\it not} impose unacceptable constraints on the curvature tensor. But whereas this is normally true as a consequence of Bianchi-type 
	identities and/or matter equations of motion, it  is true here as a consequence of the gravitational  field equation whose consistency we are checking! This is ``third-way'' consistency.
	
	A corollary is that  the EMG field equation is resistant to modification by the inclusion of additional tensor terms:  the addition of
	a generic tensor will yield equations that are inconsistent even if this additional tensor satisfies a Bianchi identity. However, there is one simple consistent 
	modification of the EMG equation and that is the addition of a parity-violating Cotton tensor
	term;  the modified field equation is
	\begin{equation}\label{EGMGeq}
	\Lambda g_{\mu\nu} + G_{\mu\nu} + \frac{1}{\mu} C_{\mu\nu} - \frac{1}{m^2}H_{\mu\nu} + \frac{1}{m^4} L_{\mu\nu} =0\, ,
	\end{equation}
	where $\mu$ is a new mass parameter, which we may take to have either sign. We shall call this ``Exotic General Massive Gravity'' (EGMG) because it is a parity-violating generalization   
	of EMG in the same sense that GMG is a parity-violating generalization of NMG. As we shall see it  also has a $4$-flavour CS-like formulation, but now with an action of
	no definite parity, although with the restriction that $\Lambda \neq - m^4/\mu^2$; in the $\mu\to\infty$ limit, this becomes the above mentioned $\Lambda\ne0$ restriction 
	on the  EMG action. 
	
	Another result of this paper is a semi-systematic construction of an infinite sequence of third-way consistent 3D gravity field equations.  The simplest example is MMG
	and the next-to-simplest example is EMG. Both are atypical  in that they admit modifications not allowed in the general case, but  the general case leads to
	equations that are higher than $4$th order.   The main advantage of the construction is that a simple modification of it  leads directly to a consistent coupling to matter. 
	This is usually trivial but a complication of third-way consistency is that the matter stress tensor is {\it not} a consistent source tensor for the metric equation \cite{Bergshoeff:2014pca}.  
	The consistent  source tensor for MMG was found  in \cite{Arvanitakis:2014yja} by making use of the CS-like action; it is quadratic in the stress tensor!  Here we recover 
	this result in a much simpler way, and extend it to EGMG. 
	
	Notice that the metric for a maximally symmetric  3D spacetime will satisfy (\ref{EGMGeq}) if $G_{\mu\nu} =-\Lambda g_{\mu\nu}$ since the Cotton tensor is identically zero 
	in such backgrounds; we may therefore identify the parameter $\Lambda$ as the cosmological constant. We analyse EMG and EGMG in a linearization about 
	an AdS vacuum ($\Lambda<0$), determining the ``no tachyon''  condition, which is always satisfied for sufficiently large AdS  radius.  We also investigate
	unitarity conditions via a linearization of the CS-like action. Because the EMG action is parity odd, one of the two spin-2 modes must be a ghost, and we confirm this.
	The same turns out to be  true for EGMG, so none of the new massive gravity models here is unitary. In addition, we show that the product of the two central 
	charges in the asymptotic  symmetry algebra is negative, so that any holographic dual CFT will certainly be non-unitary, but this is also a feature of 3D 
	conformal gravity \cite{Afshar:2011qw,Afshar:2014rwa}. 
	
	In the following we first present the new 3D ``exotic'' massive gravity models as examples arising from a systematic construction of third-way consistent field equations, 
	thereby making contact with the earlier MMG theory.  We then present their CS-like actions, a Hamiltonian analysis of them,  and results on  linearization about AdS. 
	We follow this with a systematic analysis of CS-like actions of definite parity; the results confirm that NMG and EMG are the only possibilities for propagation
	of a parity-doublet of spin-2 states if we insist on an explicit metric equation not given by an infinite series. We conclude with a discussion of our results and some comments on 
	their implications. 
	
	\section{Systematics of third-way consistency}
	\paragraph{}
	
	The new 3D massive gravity model that we have called ``Exotic Massive Gravity'' joins a very short list of  field equations that are known  to be third-way consistent; the
	only previously known examples, which are also both in 3D, are ``Minimal Massive Gravity'' \cite{Bergshoeff:2014pca}, which propagates a single spin-2 mode, and a
	modified 3D Yang-Mills  equation that is related to multi-membrane dynamics \cite{Arvanitakis:2015oga}.  Here we present a construction that yields an infinite
	sequence of  third-way consistent generalizations  of the 3D Einstein field equations. The MMG and EMG/EGMG equations constitute the simplest and next-to-simplest
	cases.
	
	Our starting point is any symmetric ``Einstein-type'' tensor  ${\mathcal{G}}_{\mu\nu}$ that satisfies the  Bianchi identity
	\begin{equation}
	\nabla^\mu {\mathcal{G}}_{\mu\nu} \equiv 0\, .
	\end{equation}
	From this we construct the following ``Schouten-type'' symmetric tensor
	\begin{equation}
	{\mathcal{S}}_{\mu\nu} =  {\mathcal{G}}_{\mu\nu} - \frac{1}{2} g_{\mu\nu} \mathcal{G} \qquad \left(\mathcal{G} = g^{\mu\nu} {\mathcal{G}}_{\mu\nu}\right),
	\end{equation}
	which  satisfies the identity
	\begin{equation}\label{ConstraintS}
	\nabla^\mu {\mathcal{S}}_{\mu\nu} \equiv \nabla_\nu {\mathcal{S}}  \, , \qquad \left({\mathcal{S}} = g^{\mu\nu} {\mathcal{S}}_{\mu\nu}\right) \, .
	\end{equation}
	Now we use the Schouten-type tensor to construct the  tensor
	\begin{equation}
	{\mathcal{H}}_{\mu\nu} =  \epsilon_\mu{}^{\rho\sigma} \nabla_\rho {\mathcal{S}}_{\nu\sigma}\, ,
	\end{equation}
	which is symmetric as a consequence of (\ref{ConstraintS}). For the choice ${\mathcal{G}}_{\mu\nu} = G_{\mu\nu}$, in which case
	${\mathcal{S}}_{\mu\nu} =S_{\mu\nu}$, this $\mathcal{H}$-tensor is the Cotton tensor, but we do {\it not} call it a ``Cotton-type'' tensor
	because it does not satisfy a Bianchi identity for any other choice of ${\mathcal{G}}_{\mu\nu}$; instead, it satisfies the identity
	\begin{equation}\label{notBianchi1}
	\nabla^\mu {\mathcal{H}}_{\mu\nu} \equiv  - \epsilon_\nu{}^{\rho\sigma} {\mathcal{S}}_\rho{}^\lambda G_{\lambda\sigma} \, .
	\end{equation}
	We shall also need the symmetric tensor
	\begin{equation}
	{\mathcal{L}}_{\mu\nu} = \frac12 \epsilon_\mu{}^{\rho\sigma}\epsilon_\nu{}^{\lambda\tau} {\mathcal{S}}_{\rho\lambda} {\mathcal{S}}_{\sigma\tau} \,,
	\end{equation}
	which satisfies the identity
	\begin{equation}\label{notBianchi2}
	\nabla^\mu {\mathcal{L}}_{\mu\nu} \equiv  - \epsilon_\nu{}^{\rho\sigma} {\mathcal{S}}_\rho{}^\lambda {\mathcal{H}}_{\lambda\sigma} \,.
	\end{equation}
	
	We now have the ingredients needed for a general construction of third-way consistent field equations, but first we consider further
	the prototypical choice ${\mathcal{G}}_{\mu\nu} \propto G_{\mu\nu}$.

	\subsection{MMG}
	
	For  ${\mathcal{G}}_{\mu\nu} = G_{\mu\nu}/\mu$ we have\footnote{The sign of the $J$ tensor is chosen to agree with  \cite{Arvanitakis:2014xna,Arvanitakis:2014yja}.} 
	\begin{equation}
	{\mathcal{H}}_{\mu\nu} = \frac{1}{\mu}C_{\mu\nu}\, , \qquad {\mathcal{L}}_{\mu\nu} = -\frac{1}{\mu^2} J_{\mu\nu}\, , \qquad 
	J_{\mu\nu} = -\frac{1}{2} \epsilon_\mu{}^{\rho\sigma}\epsilon_\nu{}^{\lambda\tau} S_{\rho\lambda} S_{\sigma\tau} \, . 
	\end{equation}
	The $J$-tensor appears in the MMG field equation
	\begin{equation} \label{MMG}
	E_{\mu\nu} \equiv \Lambda_0 g_{\mu\nu} + G_{\mu \nu}  + \frac{1}{\mu}C_{\mu\nu}  + \frac{\gamma}{\mu^2} J_{\mu\nu}  = 0 \, ,
	\end{equation}
	where $\gamma$ is an arbitrary dimensionless constant and  the constant  $\Lambda_0$ has dimensions of the cosmological constant; it {\it is} the 
	cosmological constant for $\gamma=0$, which is the TMG
	limit.  Consistency  for $\gamma\neq0$ follows from the fact that
	\begin{equation}
	\nabla^\mu E_{\mu\nu} =  -\frac{\gamma}{\mu^2}  \epsilon_\nu{}^{\rho\sigma} S_\rho{}^\lambda C_{\lambda\sigma} 
	= \frac{\gamma^2}{\mu^3} \epsilon_\nu{}^{\rho\sigma} S_\rho{}^\lambda J_{\lambda\sigma} \equiv 0\, ,
	\end{equation}
	where the second equality results from using, first,   the field equation $E_{\mu\nu}=0$ to replace the Cotton tensor, and then the symmetry of both $S$ and $S^2$.
	The final  identity also follows from symmetry of  $S$ and $S^2$, as shown in \cite{Bergshoeff:2014pca}.  The consistency of the equation $E_{\mu\nu}=0$ is
	of third-way type because it depends on the validity of this equation rather than just on Bianchi identities.

	\subsection{EMG and EGMG, and beyond}
	
	More generally, we will consider field equations of the form
	\begin{equation} \label{Generalized}
	E_{\mu\nu} \equiv \Lambda_0 g_{\mu\nu} + G_{\mu \nu}  + {\mathcal{H}}_{\mu\nu}  + {\mathcal{L}}_{\mu\nu}  = 0 \, .
	\end{equation}
	For the choice ${\mathcal{G}}_{\mu\nu} = G_{\mu\nu}/\mu$ this equation reduces to the special case of the MMG equation (\ref{MMG}) with $\gamma=1$. 
	It is a special feature of the MMG case that consistency does not determine $\gamma$; in the general case, consistency fixes the relative coefficient of the $\mathcal{H}$ and 
	$\mathcal{L}$ tensors. To prove consistency we use (\ref{notBianchi1}) and  (\ref{notBianchi2}),  and the Bianchi identity satisfied by the Einstein tensor, to deduce that 
	\begin{equation}
	\label{Consistency}
	\nabla^\mu E_{\mu\nu} =  - \epsilon_\nu{}^{\rho\sigma} {\mathcal{S}}_\rho{}^\lambda \left( G_{\lambda\sigma}  +  {\mathcal{H}}_{\lambda\sigma}  \right) 
	=  \epsilon_\nu{}^{\rho\sigma} {\mathcal{S}}_\rho{}^\lambda \left( \Lambda_0 g_{\lambda\sigma} +  {\mathcal{L}}_{\lambda\sigma} \right) \equiv 0	\, . 
	\end{equation}
	The second equality results from using the field equation $E_{\mu\nu}=0$, and the final identity uses the symmetry of the $\mathcal{S}$, $\mathcal{S}^2$ and $\mathcal{S}^3$ tensors. 
	
	Consider, for example,
	\begin{equation}
	{\mathcal{G}}_{\mu\nu} =  - \frac{1}{m^2}C_{\mu\nu} \quad \left(\Leftrightarrow \quad {\mathcal{S}}_{\mu\nu} = - \frac{1}{m^2}C_{\mu\nu} \right)\, ,
	\end{equation}
	which yields
	\begin{equation}
	{\mathcal{H}}_{\mu\nu} = - \frac{1}{m^2} H_{\mu\nu} \, , \qquad {\mathcal{L}}_{\mu\nu} = \frac{1}{m^4} L_{\mu\nu}\, , 
	\end{equation}
	where the $H$ and $L$ tensors are those of (\ref{HandL}). In this case (\ref{Generalized}) is, for  $\Lambda_0=\Lambda$,  the EMG equation (\ref{EMGeq}).
	
	This EMG example is still special in one respect: the tensor ${\mathcal{S}}_{\mu\nu}$ is traceless (because it is proportional to the Cotton tensor).
	If we add a multiple of  ${\mathcal{S}}_{\mu\nu}$  to $E_{\mu\nu}$ then the first equality of (\ref{Consistency}) is still valid but when we use $E_{\mu\nu}=0$
	in the next step we get an additional term because of the additional term in $E_{\mu\nu}$, but this additional term is proportional to
	\begin{equation}
	\epsilon_\nu{}^{\rho\sigma} {\mathcal{S}}_\rho{}^\lambda {\mathcal{S}}_{\lambda\sigma}  \, . 
	\end{equation}
	This is identically zero  for {\it any} symmetric ${\mathcal{S}}$-tensor  but this tensor will satisfy the Bianchi identity required for the validity of  the first equality of (\ref{Consistency})
	only if it is traceless, as it is for this EMG case. This special feature is what  allows us to
	modify the EMG equation to get the EGMG equation of (\ref{EGMGeq}) without sacrificing consistency.
	
	The next simplest choices for ${\mathcal{G}}_{\mu\nu}$ are found from varying, with respect to the metric, the integral
	of a curvature-squared term. For example, one could choose
	\begin{equation}
	{\mathcal{G}}_{\mu\nu} = \frac{1}{m^3} K_{\mu\nu}\, ,
	\end{equation}
	where $K_{\mu\nu}$ is the tensor appearing in the NMG equation (\ref{NMGeq}). This yields an ${\mathcal{H}}$-tensor that is
	$5$th-order in time derivatives and, ultimately, a third-way consistent field equation that is also $5$th-order in time derivatives (and presumably higher than 
	second-order in time derivatives, although we have not verified this).  An infinite number of third-way consistent field equations may be found in this way, 
	but if we restrict to equations of $4$th-order or less then the only cases are MMG and EMG/EGMG.

	\section{Matter coupling}
	
	Given a metric field equation $E_{\mu\nu}=0$, the coupling to matter is usually achieved by changing the equation to
	\begin{equation}
	E_{\mu\nu} = T_{\mu\nu} \, , 
	\end{equation}
	where $T_{\mu\nu}$ is the matter stress tensor (in some units). This modification  is not consistent 
	when the consistency of $E_{\mu\nu}=0$ is of third-way type, as it is for
	\begin{equation}\label{thirdway}
	E_{\mu\nu} \equiv \Lambda_0 g_{\mu\nu} + G_{\mu \nu}  + {\mathcal{H}}_{\mu\nu}  + {\mathcal{L}}_{\mu\nu} \, . 
	\end{equation}
	In these cases,  matter coupling can be achieved by replacing the initial ``Einstein-type'' symmetric tensor by
	\begin{equation}
	{\mathcal{G}}^\prime_{\mu\nu} \equiv  {\mathcal{G}}_{\mu\nu}  - \lambda T_{\mu\nu} \, , 
	\end{equation}
	where $\lambda$ is a constant. This still satisfies 
	\begin{equation}
	\nabla^\mu {\mathcal{G}}^\prime_{\mu\nu} =0\, , 
	\end{equation}
	but now as a consequence of the Einstein tensor Bianchi identity {\it and} the matter field equations, which imply that $\nabla^\mu T_{\mu\nu}=0$. The new Einstein-type
	tensor  gives rise to a new Schouten-type symmetric tensor ${\mathcal{S}}_{\mu\nu}^\prime$,  and two other symmetric tensors 
	$\mathcal{H}^\prime_{\mu\nu}$ and ${\mathcal{L}}^\prime_{\mu\nu}$,  defined as before: 
	\begin{equation}
	{\mathcal{H}}^\prime_{\mu\nu} =  \epsilon_\mu{}^{\rho\sigma} \nabla_\rho {\mathcal{S}}^\prime_{\nu\sigma}\,, \qquad 	
	{\mathcal{L}}_{\mu\nu}^\prime = \frac12 \epsilon_\mu{}^{\rho\sigma}\epsilon_\nu{}^{\lambda\tau} {\mathcal{S}}_{\rho\lambda}^\prime {\mathcal{S}}_{\sigma\tau}^\prime \,.
	\end{equation}
	These new tensors  satisfy the following identities
	\begin{equation}\label{prime-ids}
	\nabla^\mu {\mathcal{H}}^\prime_{\mu\nu} \equiv  - \epsilon_\nu{}^{\rho\sigma} {\mathcal{S}}^\prime_\rho{}^\lambda G_{\lambda\sigma}\,, \qquad 	
	\nabla^\mu {\mathcal{L}}'_{\mu\nu} =  - \epsilon_\nu{}^{\rho\sigma} {\mathcal{S}}^\prime_\rho{}^\lambda {\mathcal{H}}_{\lambda\sigma}^\prime \,, 
	\end{equation}
	which are entirely analogous to the identities of (\ref{notBianchi1}) and (\ref{notBianchi2}). 
	
	\subsection{MMG revisited}
	
	Recall that $\mathcal{G}_{\mu\nu} = G_{\mu\nu}/\mu$ in this case, which means that 
	\begin{equation}
	\mu\,  \mathcal{G}'_{\mu\nu} = G_{\mu\nu} - \lambda T_{\mu\nu} \equiv G'_{\mu\nu}\, , \qquad 
	\mu\,  \mathcal{S}'_{\mu\nu} = S_{\mu\nu}  - \lambda \hat T_{\mu\nu}  \equiv S'_{\mu\nu}\, , 
	\end{equation}
	where 
	\begin{equation}
	\hat T_{\mu\nu} = T_{\mu\nu} -\frac12 g_{\mu\nu} T\, . 
	\end{equation}
	As we also have $\mathcal{H}_{\mu\nu} = C_{\mu\nu}/\mu$ and $\mathcal{L}_{\mu\nu}= -J_{\mu\nu}/\mu^2$ for this MMG case,  it is convenient to use a notation
	for which 
	\begin{equation}
	\mu\, \mathcal{H}'_{\mu\nu} \equiv C^\prime_{\mu\nu} \, , \qquad \mu^2 \mathcal{L}'_{\mu\nu} \equiv  - J^\prime_{\mu\nu}\, , 
	\end{equation}
	in which case we may rewrite (\ref{prime-ids}) as 
	\begin{equation}\label{new-notation}
	\nabla^\mu C^\prime_{\mu\nu} \equiv  - \epsilon_\nu{}^{\rho\sigma}S^\prime_\rho{}^\lambda G_{\lambda\sigma}\,, \qquad 	
	\nabla^\mu J'_{\mu\nu} =  \epsilon_\nu{}^{\rho\sigma} S^\prime_\rho{}^\lambda C^\prime_{\lambda\sigma}\, . 
	\end{equation}
	
	Assuming that $\gamma\neq0$ (because otherwise we would be discussing TMG) we now replace $E_{\mu\nu}$ of (\ref{thirdway}) by
	\begin{equation} \label{MatterCoupled}
	E'_{\mu\nu} \equiv \Lambda_0 g_{\mu\nu} - \gamma^{-1} G_{\mu \nu} + \gamma^{-1}(1+\gamma\eta) G^\prime_{\mu\nu}+ 
	\frac{1}{\mu}C_{\mu\nu}^\prime  + \frac{\gamma}{\mu^2} J_{\mu\nu}^\prime   \,,
	\end{equation}
	for arbitrary constant $\eta$. Using (\ref{new-notation}) we find that 
	\begin{eqnarray}
	\nabla^\mu E'_{\mu\nu} &=&   -\frac{\gamma}{\mu} \epsilon_\nu{}^{\rho\sigma} {\mathcal{S}}'_\rho{}^\lambda \left(\gamma^{-1}G_{\lambda\sigma} 
	- \frac{1}{\mu} C'_{\lambda\sigma} \right)\nonumber \\
	&=&   -\frac{1}{\mu} \epsilon_\nu{}^{\rho\sigma} {\mathcal{S}}'_\rho{}^\lambda \left(\gamma \Lambda_0 g_{\lambda\sigma} + 
	(1+\gamma\eta) G^\prime_{\lambda\sigma} + \frac{\gamma^2}{\mu^2} J_{\lambda\sigma}^\prime \right)\, , \nonumber\\
	&=&  - \epsilon_\nu{}^{\rho\sigma} {\mathcal{S}}'_\rho{}^\lambda \left((1+\gamma\eta) \mathcal{S}^\prime_{\lambda\sigma} - 
	\frac{\gamma^2}{\mu} \mathcal{L}_{\lambda\sigma}^\prime \right) \equiv 0\, . 
	\end{eqnarray}
	We have used $E'_{\mu\nu}=0$ to get to the second line. The third line follows from the relation between the $\mathcal{G}'$ and $\mathcal{S}'$ tensors 
	and the symmetry of the latter. The final identity is, as usual, due to the symmetry of powers of the $\mathcal{S}'$ tensor. 
	
	The equation $E'_{\mu\nu}=0$ is therefore consistent. It can be written in the form 
	\begin{equation}
	\eta G_{\mu\nu}  + \Lambda_0 g_{\mu\nu} + \frac{1}{\mu} C_{\mu\nu} + \frac{\gamma}{\mu^2} J_{\mu\nu} =\Theta_{\mu\nu}\, , 
	\end{equation}
	where $\Theta_{\mu\nu}$ is the MMG source tensor. For the choice 
	\begin{equation}
	\lambda = \frac{\gamma}{(1+\gamma\eta)^2}\, , 
	\end{equation}
	the source tensor is 
	\begin{eqnarray}
	\Theta_{\mu\nu} &=& \frac{1}{(1+\gamma\eta)} T_{\mu\nu} +  \frac{\gamma}{\mu(1+ \gamma\eta)^2} \epsilon_\mu{}^{\rho\sigma}\nabla_\rho \hat T_{\nu\sigma} 
	-\frac{\gamma^2}{\mu^2(1+\gamma\eta)^2} \epsilon_\mu{}^{\rho\sigma}\epsilon_\nu{}^{\lambda\tau} S_{\rho\tau} \hat T_{\sigma\tau}  \nonumber\\
	&& + \frac{\gamma^3}{2\mu^2(1+\gamma\eta)^4}  \epsilon_\mu{}^{\rho\sigma}\epsilon_\nu{}^{\lambda\tau} \hat T_{\rho\tau} \hat T_{\sigma\tau} \, . 
	\end{eqnarray}
	This is precisely the result found by other means  in \cite{Arvanitakis:2014yja}.	
	
	\subsection{EMG/EGMG}
	
	Recall that for  EMG we have $\mathcal{G}_{\mu\nu} = - C_{\mu\nu}/m^2$, so that 
	\begin{equation}
	- m^2\,  \mathcal{G}'_{\mu\nu} =  C_{\mu\nu} - \lambda T_{\mu\nu} 
	\end{equation}
	and hence 
	\begin{equation}
	- m^2\,  \mathcal{S}'_{\mu\nu} =  C_{\mu\nu} - \lambda \hat T_{\mu\nu} \equiv C'_{\mu\nu}\, ,  
	\end{equation}
	which then gives us 
	\begin{equation}
	- m^2 \mathcal{H}'_{\mu\nu} =  \epsilon_\mu{}^{\rho\sigma} C'_{\nu\sigma}\, , \qquad 
	m^4\mathcal{L}'_{\mu\nu} = \frac{1}{2}\epsilon_\mu{}^{\rho\sigma} \epsilon_\nu{}^{\lambda\tau} C'_{\nu\sigma} C'_{\sigma\tau}\, . 
	\end{equation}
	A peculiarity of this EMG case is that the  modified Einstein-type tensor 
	has no definite parity, which implies that the source for the EMG field equation will break parity.  This suggests that we should consider 
	matter coupling to EMG  in the context of its parity-violating EGMG extension, which motivates us to replace $E_{\mu\nu}$ of (\ref{thirdway}) by
	\begin{equation}
	E'_{\mu\nu} \equiv \Lambda_0 g_{\mu\nu} + G_{\mu \nu}  - \frac{m^2}{\mu} \mathcal{G}'_{\mu\nu} + {\mathcal{H}}'_{\mu\nu}  + {\mathcal{L}}'_{\mu\nu} \, .  
	\end{equation}
	For $\lambda=0$ the equation $E'_{\mu\nu}=0$ is the EGMG equation (\ref{EGMGeq}). 
	To prove the consistency  for $\lambda\neq0$ we use the identities  (\ref{prime-ids}) to compute
	\begin{eqnarray}
	\nabla^\mu E'_{\mu\nu} &=&  - \epsilon_\nu{}^{\rho\sigma} {\mathcal{S}}'_\rho{}^\lambda \left(G_{\lambda\sigma}  +  {\mathcal{H}}'_{\lambda\sigma}  \right) \nonumber \\
	&=&  \epsilon_\nu{}^{\rho\sigma} {\mathcal{S}}'_\rho{}^\lambda \left( \Lambda_0 g_{\lambda\sigma} -  \frac{m^2}{\mu} \mathcal{G}'_{\lambda\sigma} + {\mathcal{L}}'_{\lambda\sigma} \right)\nonumber\\
	&=&   \epsilon_\nu{}^{\rho\sigma} {\mathcal{S}}'_\rho{}^\lambda \left( - \frac{m^2}{\mu} \mathcal{S}'_{\lambda\sigma} + {\mathcal{L}}'_{\lambda\sigma} \right) \equiv 0\, . 
	\end{eqnarray}
	We have  used $E'_{\mu\nu}=0$ to arrive at the second line. The third line follows from the relation between the $\mathcal{G}'$ and $\mathcal{S}'$ tensors and the symmetry of latter; the 
	final identity is due to the symmetry of powers of the $\mathcal{S}'$ tensor. 
	
	We may write the equation $E'_{\mu\nu} =0$ in the form 
	\begin{equation}
	G_{\mu\nu} + \lambda_0 g_{\mu\nu} + \frac{1}{\mu} C_{\mu\nu} - \frac{1}{m^2} H_{\mu\nu} + \frac{1}{m^4} L_{\mu\nu} = \Theta_{\mu\nu}\, , 
	\end{equation}
	where the EGMG source tensor is 
	\begin{equation}
	\Theta_{\mu\nu}=  \frac{\lambda}{\mu} \hat T_{\mu\nu} - \frac{\lambda}{m^2} \epsilon_\mu{}^{\rho\sigma}\nabla_\rho \hat T_{\nu\sigma} + 
	\frac{2\lambda}{m^4} \epsilon_\mu{}^{\rho\sigma} \epsilon_\nu{}^{\lambda\tau} 
	C_{\rho\lambda}\hat T_{\sigma\tau} - \frac{\lambda^2}{m^4} \epsilon_\mu{}^{\rho\sigma} \epsilon_\nu{}^{\lambda\tau}  \hat T_{\rho\lambda} \hat T_{\sigma\tau}\, . 
	\end{equation}
	Notice that 
	\begin{equation}
	\lambda = \mu \quad \Rightarrow\quad \Theta_{\mu\nu} = T_{\mu\nu}  + {\cal O} \left(\mu/m^2\right)\, , 
	\end{equation}
	which becomes the standard source term for TMG in the $m^2\to\infty$ limit, but for finite $m^2$ the $\mu\to\infty$ limit is no longer
	possible (because it implies $\lambda\to\infty$). 
	
	The $\mu\to\infty$ limit {\it is}  possible for the choice $\lambda=m$, and in this case one has the matter-coupled EMG equation 
	\begin{equation}
	G_{\mu\nu} + \Lambda_0 g_{\mu\nu} - \frac{1}{m^2} H_{\mu\nu} + \frac{1}{m^4} L_{\mu\nu} = \frac{1}{m}\epsilon_\mu{}^{\rho\sigma} \nabla_\rho \hat T_{\nu\sigma} + {\cal O}\left( 1/m^3\right) . 
	\end{equation}
	This is indeed exotic but not surprising in light of our earlier observation that coupling matter to EMG breaks parity!

	\section{CS-like and Hamiltonian formulations}
	
         The EMG and NMG equations have the same linearized limit in an expansion about a Minkowski vacuum, and similarly for EGMG and GMG.
	This tells us that EMG propagates a parity doublet of massive spin-2 modes in this vacuum, and that EGMG propagates these spin-2 modes but with the
	mass degeneracy lifted by parity violation.  This implies that the physical phase space of {\it linearized}  EMG and EGMG has dimension
	$4$ per space point, but it is far from clear whether this is also true of  the full non-linear equations.   This issue is most easily addressed
	in the context of a Chern-Simons-like formulation since it is then a short step to the Hamiltonian formulation, which allows a simple background-independent
	determination of the physical phase-space dimension.
	
	The general  $N$-flavour CS-like model is defined by a Lagrangian 3-form constructed from a set $\{a^r; r=1, \dots,N\}$ of Lorentz-vector valued
	one form fields by exterior multiplication, without the use of a metric (which is implicit in the identification of one member of the set as an invertible dreibein).	
	Making use of a dot and cross product notation for 3D Lorentz vectors, the general Lagrangian 3-form of this type may be written as
	\begin{equation}\label{CSlike}
	L = \frac12 g_{rs} a^r \cdot da^s + \frac16 f_{rst} a^r \cdot a^s \times a^t \,,
	\end{equation}
	where the exterior product of forms is implicit. The coupling constants $g_{rs}$ and $f_{rst}$ can be viewed as symmetric tensors of an $N$-dimensional
	``flavour space'', with $g_{rs}$ a metric  for this space if we assume it be invertible. These coupling constants are restricted by the requirement that the $a^r$
	include the dreibein $e$ and the dual spin-connection $\omega$, such that $L$ is invariant (or transforms into a closed $3$-form) under local Lorentz
	transformations.  The  local-Lorentz covariant extensions of $de$ and $d\omega$ are the torsion and curvature two-forms: respectively, 
	\begin{equation}
	T(\omega) \equiv D(\omega)e = de+ \omega\times e\, , \qquad
	R(\omega) = d\omega + \frac{1}{2} \omega\times\omega\, . 
	\end{equation}
	The only way that $\omega$ can otherwise appear is through the covariant derivative $D(\omega)$ or the 
	Lorenz-Chern-Simons 3-form 
	\begin{equation}
	L_{LCS} = \frac{1}{2} \left(\omega \cdot d\omega + \frac{1}{3}\omega\cdot\omega\times\omega\right)\, . 
	\end{equation}
	
	For $N=2$ the local Lorentz invariance  is sufficient to imply that $L$ is a CS action for a 3D gravity model with no local degrees of freedom.
	For $N=3$ there is one additional 1-form field, which can be used to construct a CS-like action for the parity-violating massive gravity theories
	TMG and MMG. These propagate a single spin-2 mode in a Minkowski vacuum  and, more generally, have a physical phase space whose
	dimension per space point is $2$. The $N=4$ CS-like theories generically have a phase space whose dimension per space point is $4$, implying the propagation of two modes.
	Both NMG  and ZDG can be formulated as $N=4$ CS-like theories, as can their parity-violating extensions, and these
	all  have the expected physical phase space  dimension. This result is  consistent with the fact that two massive spin-2 modes are propagated
	in a Minkowski vacuum, and it also tells us that no new local degrees of freedom can appear in any other background.
	We now aim to establish the same result for EMG and EGMG, and the first step is to find a CS-like formulation for them.
	
	We undertake a more systematic analysis of $N=4$ CS-like models in the following section. It will be shown there that the attempt to find an 
	action of this type for the EMG equations (\ref{EMGeq}) leads uniquely to the following Lagrangian three-form constructed from 
	one-form fields $\{e,\omega,h,f)$:
	\begin{eqnarray}\label{EMG-CSL}
	L_{EMG} &=&  f \cdot R(\omega)  + \frac{1}{6 m^4} f \cdot f \times f -  \frac{1}{2m^2} f \cdot D(\omega) f - \frac{\Lambda}{2} f \cdot e \times e  \nonumber \\
	&-& m^2 h \cdot T(\omega) - (m^2+\Lambda) L_{LCS} \, . 
	\end{eqnarray}	
	The auxiliary field $h$ is parity even and has dimensions of mass-squared,  while $f$ is parity odd
	and has dimensions of mass-cubed.  Integrating $L$ over a 3-manifold  with local coordinates $x^\mu$ ($\mu=0,1,2$),  we find that 
	\begin{equation}
	I_{EMG}[e,\omega,f,h] \ \propto \ \int_M L_{EMG} = \frac{1}{6} \int\! d^3\!x \, \varepsilon^{\mu\nu\rho} L_{\mu\nu\rho}\, , 
	\end{equation}
	where the constant of proportionality  has dimensions of inverse mass-squared  in units for which $\hbar=c=1$, and $L_{\mu\nu\rho}$ are the components of $L_{EMG}$. 
	The remarkable feature of this result is that $I_{EMG}$ is parity-odd.  This  is sufficient for the field equations to preserve parity, even though
	the action is not parity invariant; this is what makes EMG `exotic'.  In principle, the fact that we have found a parity-odd CS-like action for EMG does not preclude the 
	existence of some   other parity-even action, but another result  of our later systematic analysis  is that {\it there is no even-parity $N=4$ CS-like action for EMG}. 	
	
	A slight modification of the EMG Lagrangian 3-form is sufficient to describe EGMG:
	\begin{eqnarray}\label{EGMGAction1}
	L_{EGMG} &=&  f \cdot R(\omega)  + \frac{1}{6 m^4} f \cdot f \times f -  \frac{1}{2m^2} f \cdot D(\omega) f + \frac{\nu}{2} f \cdot e \times e  \\
	&-& m^2 h \cdot T(\omega) + (\nu - m^2) L_{LCS}  + \frac13 \frac{\nu m^4}{\mu} e \cdot e \times e \nonumber\, ,
	\end{eqnarray}	
	where
	\begin{equation}\label{nutoL}
	\nu =  -\Lambda - \frac{m^4}{\mu^2}\, .
	\end{equation}
	Apart from the $\mu$-dependent  modification of some of the coefficients of $L_{EMG}$, there is one additional term, proportional to $e\cdot e\times e$, that
	was absent from $L_{EMG}$, and it leads to a parity-even term in the action; paradoxically,  this term is responsible for the parity violation of the  EGMG field equations.
	
	We shall now focus on the EGMG case since EMG is the subcase with $|\mu|=\infty$. The equations of motion obtained from  $L_{EGMG}$ by variation 
	with  respect to $(e,\omega,h,f)$ are	
	\begin{eqnarray}\label{fieldeqs}
	\delta e\, : \qquad 0&=& Dh - \frac{\nu}{m^2} e\times f - \frac{2\nu m^2}{\mu} e\times e \nonumber\\
	\delta \omega\, : \qquad 0&=& Df + (\nu-m^2) R - \frac{1}{2m^2} f\times f - m^2 e\times h \nonumber \\
	\delta h\, :\qquad 0&=& T(\omega) \nonumber \\
	\delta f\, :\qquad 0&=& R - \frac{1}{m^2} Df + \frac{1}{2m^4} f\times f + \frac{\nu}{2} e\times e\, .
	\end{eqnarray}	
	Integrability of these  2-form equations imposes the following 3-form  conditions 
	\begin{equation}
	e (e\cdot h) = 0 \, , \qquad e(e\cdot f) = 0\, , \qquad m^6 h (e\cdot h) + \nu^2 f(e\cdot f)=0\, .
	\end{equation}
	For an invertible dreibein $e$, this requires
	\begin{equation}\label{secondary}
	e\cdot h =0\, , \qquad e\cdot f =0\, .
	\end{equation}
	These 2-form equations are relevant to the Hamiltonian formulation, to be discussed below,  since the space-space components
	are constraints on canonical variables.
		
	The field equations obtained from variation of $(\omega,h,f)$ are jointly equivalent to
	\begin{equation}
	T(\omega) =0\, , \qquad e\times h = \frac{\nu}{m^2}\left[R + \frac{m^2}{2} e\times e\right]\, ,
	\end{equation}
	which may be solved algebraically for $\omega$ and $h$, given invertibility of the dreibein, and the one further equation
	\begin{equation}\label{further-eq}
	0= \frac{\nu}{2} e\times e + R(\omega) - \frac{1}{m^2} D(\omega)f +  \frac{1}{2m^4} f\times f\, ,
	\end{equation}
	which {\it cannot} be solved algebraically for $f$.  However, it is possible to solve algebraically for $f$ from the
	equation obtained from variation of the dreibein $e$, as may be seen by writing this equation in the form
	\begin{equation}
	e\times f =  \frac{m^2}{\nu} D(\omega)h - \frac{2m^4}{\mu} e\times e\, .
	\end{equation}
	Given that we have already solved for $h$ in terms of $e$ and $\omega$, we may now solve this equation algebraically
	for $f$ in terms of $e$ and $\omega$. By solving the zero torsion equation for $\omega$ in terms of $e$, in the usual way,
	we thereby  have  explicit expressions for $h$ and $f$ in terms of the curvature $R$ and its covariant derivatives.
	Using these in (\ref{further-eq}) yields a field equation for the dreibein $e$. 
	
	By introducing the metric and auxiliary tensors 
	\begin{equation}
	g_{\mu\nu} \equiv  e_\mu\cdot e_\nu\,, \qquad h_{\mu\nu} \equiv e_\mu \cdot h_\nu \, , \qquad f_{\mu\nu} \equiv e_\mu \cdot f_\nu\, , 
	\end{equation}
	we may express the results of solving for the auxiliary one-forms $h$ and $f$ in the following tensor form:
	\begin{equation}\label{hf-solutions}
	h_{\mu\nu} =  \frac{\nu}{m^2} S_{\mu\nu}+ \frac{\nu}{2}g_{\mu\nu} \, , \qquad 
	f_{\mu\nu} =  C_{\mu\nu} + \frac{m^4}{\mu} g_{\mu\nu}\, .
	\end{equation}
	Notice that these tensors are symmetric, as required by (\ref{secondary}).
	On substituting these results into the equation (\ref{further-eq}) one recovers the EGMG field equation (\ref{EGMGeq}).
	This confirms our claim that the field equations of the action (\ref{EGMGAction1}) are equivalent to the EGMG equations,
	assuming invertibility of the dreibein,  and the same follows for EMG by taking the $|\mu|\to\infty$ limit.
	
	It is important to appreciate here that we may not use (\ref{hf-solutions}) to eliminate $h$ and $f$ from the action
	to get an equivalent action for $e$ alone. This is because the solution for $f$ required the use of the $e$ equation.
	Thus, the existence of an action implies consistency of the equations for the metric but this consistency is necessarily of third-way
	type because there is no action functional of the metric alone that yields these equations.

	\subsection{Hamiltonian formulation}
	
	Now we make use of a general procedure for passing from a CS-like action (\ref{CSlike}) to  a Hamiltonian formulation. 		
	For CS theories one simply has to
	rewrite the CS action by performing a time-space split: $\mu = (0,i)$ with $i=1,2$, so that
	\begin{equation}
	a^r = a^r_0 dt + a_i^r dx^i\, .
	\end{equation}
	Substitution into (\ref{CSlike}) yields
	\begin{equation}
	L= - \frac12 \varepsilon^{ij} g_{rs} a_i^r \cdot a_j^s + a_0^r \cdot \phi_r \,,
	\end{equation}
	where $\varepsilon^{ij} \equiv \varepsilon^{0ij}$ and
	\begin{equation}
	\phi_r = \varepsilon^{ij} \left( g_{rs} \partial_i a_{j}^s + \frac12 f_{rst} a_i^s \times a_j^t \right) \,.
	\end{equation}
	The time components of the one-form fields are now Lagrange multipliers for the $N$ Lorentz-vector primary constraints $\phi_r=0$. For CS
	theories these $N$ constraints form a first class set and hence generate $N$ gauge invariances, sufficient to ensure that the dimension per space point
	of the physical phase space is zero; i.e. there are no local degrees of freedom.
	
	For the more general CS-like models,  the count of degrees of freedom is different for two reasons. Firstly, not all $N$ primary constraints are first class and, secondly,
	one must take into account ``secondary'' constraints  arising from the assumed invertibility of the dreibein.  These are the space-space component of
	2-form equations (\ref{secondary}), i.e.
	\begin{equation}
	0= \varepsilon^{ij} e_i \cdot h_j \equiv \Delta^{eh} \, , \qquad 0= \varepsilon^{ij} e_i \cdot f_j \equiv \Delta^{ef}\, . 
	\end{equation}
	Without the invertibility assumption for $e$,  we could 
	interpret the equations (\ref{secondary}) as constraints on the Lagrange multipliers $(e_0,h_0,f_0)$, in accordance with
	Dirac's prescription for construction of the Hamiltonian \cite{Dirac:1950pj}. The above constraints are therefore not  ``secondary''  in Dirac's sense
	and must be dealt with differently \cite{Bergshoeff:2014bia,Hajihashemi:2017svz}.  Here we follow the 
	procedure of  \cite{Bergshoeff:2014bia} in which these constraints are omitted from the ``total Hamiltonian''; consistency then requires 
	certain conditions on the Poisson bracket relations of the `secondary' constraints, but we find that these are satisfied.

	To proceed, it  is convenient to first integrate the primary constraint functions, over a spacelike hypersurface $\Sigma$, against a set of
	smooth Lorentz-vector valued  fields $\{\xi^r ; r=1,\dots,N\}$. We then have a basis for the (infinite dimensional) vector space of primary constraints provided by
	functionals of the form
	\begin{equation}
	\varphi [\xi] = \int_\Sigma d^2 x \, \xi^r_a \phi_r^a + Q[\xi] \,,
	\end{equation}
	where $Q[\xi]$ is a boundary term that we add, in the case that $\Sigma$ has a boundary,  to ensure that  these functionals have
	well-defined functional derivatives \cite{Merbis:2014vja}.
	The Poisson bracket of two such functionals, corresponding to fields $\xi$ and $\eta$,  is
	\begin{eqnarray}
	\{\varphi[\xi], \varphi[\eta]  \}_{PB} &=& \varphi[[\xi,\eta]] + \int_\Sigma d^2 x \, \xi_a^r \, \eta_b^s \, {\mathcal{P}}_{rs}^{ab} \nonumber \\
	&&- \,  \int_{\partial \Sigma} dx^i \, \xi^r \cdot (g_{rs} \partial_i \eta^s + f_{rst}\, a^s_i \times \eta^t) \,,
	\end{eqnarray}
	where $[\xi,\eta]^t = f^t{}_{rs} \xi^r \times \eta^s$, and
	\begin{equation}
	{\mathcal{P}}_{rs}^{ab} = f^t{}_{q[r} f_{s]pt} \eta^{ab} \Delta^{pq} + 2 f^t{}_{r[s} f_{q]pt} (V^{ab})^{pq} \, ,
	\end{equation}
	with
	\begin{equation}
	V_{ab}^{pq} = \varepsilon^{ij} a_{i\, a}^p a_{j\, b}^q\,, \qquad \Delta^{pq} = \varepsilon^{ij} a_i^p \cdot a_j^q \,.
	\end{equation}
	
	For the particular case of relevance here, we shall need to use the fact that
	the 3-form (\ref{EGMGAction1}) is the special  $N=4$ case of the general CS-like Lagrangian 3-form (\ref{CSlike}) with
	\begin{eqnarray}
	&& g_{f\omega} = 1 \,,\quad g_{ff} = -\frac1{m^2} \,, \quad g_{\omega\omega} =  (-m^2 + \nu) \,, \quad g_{he} = -m^2 \,,\nonumber\\
	&& f_{f\omega\omega} = 1 \,, \quad f_{fee} = \nu  \,, \quad f_{ff\omega} = -\frac1{m^2} \,, \quad f_{fff} = \frac{1}{m^4} \,, \nonumber\\
	&& f_{\omega\omega\omega} =  (-m^2 + \nu)\,, \quad f_{eh\omega} = -m^2 \,, \quad f_{eee} = \frac{2\nu m^4}{\mu}\,,
	\end{eqnarray}
	where we use the names  of the four one-form fields as labels replacing $r=1,2,3,4$. 	
	
	As we have seen, we have two `secondary'  constraints in this case, which are $\Delta^{eh}=\Delta^{ef}=0$. Their Poisson bracket
	is proportional to $\Delta^{ef}$ and hence zero on the surface defined by the enlarged set of constraints, but their Poisson brackets
	with the primary constraints is non-zero, such that the rank of the full matrix of Poisson brackets of constraint functions is the rank of
	the sub-matrix $({\mathcal{P}}_{ab})_{rs}$, but evaluated on the surface defined by the full set of constraints. We find that
	\begin{equation}
	({\mathcal{P}}_{ab})_{rs} = \left(\begin{array}{cccc}
		0 &\,  0 & \,  0&\, 0 \\
		0 &\, -\frac{\nu}{m^2} V^{ff}_{ab} - \frac{m^4}{\nu} V^{hh}_{ab} &\, \frac{m^4}{\nu}  V^{he}_{ab} &\,  \frac{\nu}{m^2} V^{fe}_{ab} \\
		0&\, \frac{m^4}{\nu}  V^{eh}_{ab}  &\, -\frac{m^4}{\nu}  V^{ee}_{ab} &\, 0 \\
		0  &\,  \frac{\nu}{m^2} V^{ef}_{ab} &\, 0 &\,  \frac{m^4}{\nu}  V^{ee}_{ab}
	\end{array}\right)
	\end{equation}
	The rank of this matrix is 4, which is therefore the dimension of the physical phase space. This result is expected from
	the fact that the linearized theory propagates two massive (spin-2) modes, and this allows us to conclude that no additional  modes appear
	in the non-linear theory.
	
	\section{Linearization about AdS}
	
	Any maximally symmetric solution of the EGMG equation  (\ref{EGMGeq}) also solves the simpler equation
	\begin{equation}
	G_{\mu\nu} = -\Lambda g_{\mu\nu}\, . 
	\end{equation}
        The parameter $\Lambda$ is therefore the cosmological constant; its value determines whether the vacuum is Minkowski,  de Sitter (dS) or anti de Sitter (AdS) according to whether 
        $\Lambda$ is zero, positive or negative. This is in contrast to NMG/GMG, for which the cosmological constant is a quadratic function of a cosmological parameter $\Lambda_0$. 
        
        In  the CS-like formulation, a maximally symmetric vacuum solution has 
        \begin{equation}
        e= \bar e \, , \qquad \omega = \bar\omega, 
        \end{equation}
        such that 
          \begin{equation}
       \bar T \equiv D(\bar\omega) \bar e = 0\, , \qquad \bar R \equiv R(\bar\omega) = \frac{1}{2} \Lambda \, \bar e\times \bar e\, . 
       \end{equation}
       In addition, the auxiliary fields take the form
        \begin{equation}
        \qquad h= C_h \bar e\, , \qquad f= C_f \bar e\, , 
        \end{equation}
       for constants $C_h$ and $C_f$.  From the field equations (\ref{fieldeqs}), and recalling the relation (\ref{nutoL}) between the parameters $\nu$  and $\Lambda$, we  learn that 
       \begin{equation} 
       C_h = -\frac{1}{2}\left(\Lambda + \frac{m^4}{\mu^2}\right)\left(1+ \frac{\Lambda}{m^2}\right)\, , \qquad C_f = - \frac{m^4}{\mu}\, .  
       \end{equation}
           	
	Following \cite{Bergshoeff:2014pca}, we now expand the 1-form fields about this vacuum solution by writing
	\begin{eqnarray}
	e = \bar{e} + k \,, && \quad h = - \left(\Lambda+ \frac{m^2}{\mu^2}\right)\left[ \frac{1}{2} \left(1+ \frac{\Lambda}{m^2}\right) + p \right] \, ,  \nonumber\\
	\omega = \bar{\omega} + v \,, && \quad  f = - \frac{m^4}{\mu}(\bar{e} + k) + q \,,
	\end{eqnarray}
	where $(k,v,p,q)$ are perturbations.  Substitution into the field equations (\ref{fieldeqs}) yields the linearized equations
	\begin{eqnarray}\label{linearEqs}
	0 &=& \bar{D} v - \Lambda \bar e \times k  - \frac{1}{m^2} \bar{D} q  - \frac{1}{\mu} \bar e \times q  \,,\nonumber\\
	0 &=&  \bar{D} v - \Lambda  \bar e \times k - m^2 \bar e \times p \,,\nonumber\\
	0 &=& \bar{D} k + \bar e \times v \,,\nonumber\\
	0 &=&  m^2 \bar{D} p -  \bar e \times q \,,
	\end{eqnarray}
	where $\bar D= D(\bar\omega)$. 
	
	We will not consider here the further analysis of these equations for the dS vacuum with $\Lambda>0$. Instead we
	proceed by supposing that $\Lambda \le0$, so that 
	\begin{equation}
	\Lambda = - 1/\ell^2\, , 
	\end{equation}
	where $\ell$ is the adS radius of curvature. The Minkowski vacuum ($\Lambda=0$)  is found by taking the $\ell\to\infty$ limit.

	\subsection{No-tachyon conditions}	
	
	In order to diagonalize the linear equations about an AdS vacuum, we set
	\begin{eqnarray}
	q &=& \frac{m^4}{2\mu^2}  \left(\sqrt{1 + \frac{4\mu^2}{m^2}}\right) (\phi_+ - \phi_-) + \left( \frac{m^4 \ell^2 + 2 m^2 \mu^2 \ell^2 - 2 \mu^2}{2\mu^2 \ell^2} \right) (\phi_+ +\phi_- ) \,,\nonumber\\
	p &=& \left(\frac{1 + m^2 \ell^2}{2\mu m^2\ell^2}\right) (\phi_+ + \phi_-) + \left(\frac{ m^2 \ell^2 - 1}{2\mu m^2\ell^2}\right)\left( \sqrt{1+\frac{4\mu^2}{m^2}}\right) (\phi_+ - \phi_-) \,,\nonumber\\
	k &=& \frac{1}{2\mu } \left[ \left(\sqrt{1 + \frac{4 \mu^2}{m^2}}\right) (\phi_+ - \phi_-) -   (\phi_+ + \phi_-) \right] - \ell f_+ + \ell f_- \,,\nonumber\\
	v &=& \phi_+ + \phi_- + f_+ + f_- \,, 
	\end{eqnarray}
        where $(\phi_+,\phi_-, f_+,f_-)$ is a new basis for the perturbation one-forms.  The field equations (\ref{linearEqs}) in this basis are
        \begin{eqnarray}
	0 &=& \bar{D} \phi_+ + M_+ \bar{e} \times \phi_+ \,,\nonumber\\
	0 &=& \bar{D} \phi_- - M_- \bar{e} \times \phi_-  \,,\nonumber\\
	0 &=& \bar{D} f_+ + \ell^{-1} \, \bar e \times f_+ \,,\nonumber\\
	0 &=& \bar{D} f_- - \ell^{-1}\,  \bar e \times f_- \,,
	\end{eqnarray}
	where 
	\begin{equation}
	M_\pm  =  m\left[\sqrt{1+ \frac{m^2}{4\mu^2}} \pm \frac{m}{2\mu}\right] \, , 
	\end{equation}
	which gives $M_\pm =m$ in the EMG ($\mu\to\infty$) limit. 
	
	This result for $M_\pm$ is independent of $\ell$ and hence applies in the Minkowski limit, for which $M_\pm$ are the 
	masses of the two propagating modes, of helicities $\pm2$.  As expected, the masses $M_\pm$ agree with those found for GMG
	in a Minkowski background. 
	
	In a background with $\Lambda\ne0$, the particle masses are $\mathcal{M}_\pm$, where $\mathcal{M}_\pm^2 =M_\pm^2 +\Lambda$, and  the no-tachyon condition is (for a mode of non-zero spin)
	$\mathcal{M}_\pm^2>0$. Equivalently $(\ell M)_\pm^2> 1$, which is itself equivalent to\footnote{Equality implies logarithmic modes, which we ignore here; 
	see \cite{Bergshoeff:2010iy} for a review of the AdS$_3$ case.}
	\begin{equation}
	( m\ell)^2-\frac{m^2\ell}{|\mu|}-1>0 \, . 
	\end{equation}
	This requires
	\begin{equation}\label{notach}
	m\ell > \sqrt{1+ \frac{m^2}{4\mu^2}} + \frac{m}{2|\mu|}\, . 
	\end{equation}
	For EMG this no-tachyon condition reduces to $m\ell >1$, which requires the AdS radius to be larger than the scale set by $1/m$. For $\mu\ne0$,  this lower 
	bound becomes more restrictive but, for any given ratio 
	$m/|\mu|$, it is satisfied for sufficiently large AdS radius.

\subsection{Unitarity} 

         When the EGMG Lagrangian 3-form (\ref{EGMGAction1}) is expanded about the AdS vacuum to second order in perturbations, 
         the second-order term $L^{(2)}$  takes the following form in the basis $(\phi_+,\phi_-, f_+,f_-)$:
	\begin{eqnarray}\label{quadact}
	2L^{(2)} &=&  \ell a_-  f_-(df_- - \ell^{-1} \, \bar e \times f_-) - \ell a_+ f_+(df_+ + \ell^{-1} \, \bar e \times f_+) \nonumber\\
	&+&\! \frac{b_-}{M_-} \phi_-(d\phi_- - M_- \bar{e} \times \phi_-) - \frac{b_+}{M_+} \phi_+(d\phi_+ + M_+ \bar{e} \times \phi_+)\, , 
	\end{eqnarray}
	where 
	\begin{equation}\label{aandb1}
	a_\pm = \frac{m}{\mu} \mp  \frac{1}{\ell m} \left(1- \frac{\nu}{m^2}\right) \, , \qquad \nu = \frac{1}{\ell^2} - \frac{m^4}{\mu^2}\, , 
	\end{equation}
	and 
	\begin{equation}\label{aandb2}
	b_\pm = \frac{mK}{\mu} \pm \sqrt{\left(\frac{mK}{\mu}\right)^2 - K  a_+a_-}\, , \qquad K= \frac{\nu\left(m^2+4\mu^2\right)}{4m^2\mu^2}\, .
	\end{equation}
	The relation of the parameter $\nu$  to the AdS radius is just the formula (\ref{nutoL}) for the AdS case. Notice that 
	\begin{equation}\label{b2a}
	b_+ b_- =  Ka_+a_-\, , 
	\end{equation}
	and that 
        \begin{equation}\label{a+a-}
         a_+a_- = \frac{m^2}{\mu^2} - \frac{1}{(m\ell)^2}\left(1- \frac{\nu}{m^2}\right)^2\, . 
        \end{equation}

        Following the similar analysis for MMG in \cite{Bergshoeff:2014pca},  we conclude from the form of $L^{(2)}$ that the no-ghost conditions for perturbative unitarity are
	\begin{equation}
	b_+>0  \, , \qquad b_->0\, . 
	\end{equation}
	These conditions imply that $b_+b_->0$, which is certainly {\it not} satisfied in the EMG limit for which $|\mu|\to\infty$; in this case
	\begin{equation}\label{EMG-factors}
	K= (\ell m)^{-2} \, , \quad a_\pm = \mp (\ell m)^{-3}\left[1+ (\ell m)^2\right]  \qquad (EMG)
	\end{equation}
	and hence $b_+b_-<0$. We conclude that EMG is not perturbatively unitary.  This was to be expected because a parity-odd action for a parity doublet implies that one
	of the two modes is a ghost; a simple spin-1 example can be found in \cite{Bergshoeff:2009tb}. 

	A necessary condition for  non-perturbative unitarity in AdS is that the asymptotic 
	Virasoro $\oplus$ Virasoro symetry algebra implied by standard  Brown-Henneaux boundary conditions have positive central charges $c_\pm$. 
	These central charges are easily 
	determined in the CS-like  formalism \cite{Merbis:2014vja};  one finds that  $c_\pm \propto a_\pm$  for a positive constant of proportionality that 
	depends on the normalization of the action. 
	Thus, positivity of the central charges is equivalent to the conditions
	\begin{equation}
	a_+ >0\, \qquad a_- >0\, .  
	\end{equation}
	This requires $a_+a_->0$, so it is already clear from (\ref{b2a}) that the conditions of perturbative unitarity in AdS and positive central charges for the asymptotic symmetry algebra 
	cannot {\it both} be satisfied when $K<0$, which is equivalent to $\nu<0$ . In fact, when $\nu<0$ we see from (\ref{aandb1}) that 
	\begin{equation}
	a_+a_- > \frac{m^2}{\mu^2} - \frac{1}{(m\ell)^2} = -\nu/m^2 >0\, ,  \qquad (\nu<0)
        \end{equation} 
	so that $b_+b_-<0$ and perturbative unitary is not possible.
	
	That leaves $K>0$, which is equivalent to $\nu>0$ (recall that no CS-like EGMG action exists for $\nu=0$). To analyse this case, we rewrite the expression (\ref{a+a-}) in the form
	\begin{eqnarray}
	a_+a_- =  -\frac{1}{(m\ell)^4}\, \left(\frac{\nu}{m^2}\right)\left[(m\ell)^2 -1 - \frac{m^2\ell}{|\mu|}\right] \left[(m\ell)^2 -1 + \frac{m^2\ell}{|\mu|}\right] \, . 
	\end{eqnarray}
        The no-tachyon condition (\ref{notach}) implies that both bracketed expressions are positive, so that $a_+a_-<0$, and hence $b_+b_-<0$,  for $\nu>0$.

        To conclude, the CS-like action for EMG/EGMG does not yield even a perturbatively unitary theory. This is disappointing but certainly 
        no surprise for EMG because of its parity odd action.

	\section{Systematics of CS-like actions}

	The most general four-flavour ($N=4$) CS-like action can be written as	
	\begin{eqnarray}
	\label{ewhf}
	L &=& a_{1} e \cdot R (\omega) + \frac{1}{3} a_2 e \cdot e \times e + a_3 \, e \cdot f \times f + a_4 \, e \cdot e \times f \nonumber\\
	&& + a_5 \, e \cdot h \times h + a_6 e \cdot e \times h  + a_7 \, e \cdot f \times h \nonumber\\
	&& + a_8 \, f \cdot R(\omega) + a_9 \, f \cdot T(\omega) + a_{10} \, f \cdot f \times f + a_{11} \,f\cdot h \times h+ a_{12}\, f \cdot f \times h  \nonumber\\
	&& + a_{13} \, f \cdot D(\omega) f + a_{14} \, f \cdot D(\omega) h \nonumber\\
	&& + a_{15} \, h \cdot R(\omega) + a_{16} \, h \cdot T(\omega) + a_{17} \, h \cdot h \times h + a_{18} \, h \cdot D(\omega) h \nonumber\\
	&& + a_{19} \Big(\omega \cdot d\omega + \frac13 \omega \cdot \omega \times \omega \Big) + a_{20} e \cdot T (\omega) \,.
	\end{eqnarray}
	We restrict our attention to this $N=4$ case as $N>4$ leads to higher than $4$th order  metric equations that propagate at least one spin-2 mode that is 
	either a tachyon or a ghost \cite{Afshar:2014ffa}, and the $N<4$  possibilities are already known. As mentioned in the introduction, our aim will be 
	to identify those cases for which the field equations allow $h$ and $f$ 
	(and $\omega$) to be eliminated {\it algebraically and finitely}, i.e. such that the result is not given by an infinite
	series of terms. This will exclude theories such as ZDG, but will include NMG and GMG, possibly in more than one way, in addition to the EMG and EGMG theories
	presented in the previous sections. The main issue is whether there are any additional possibilities. 	
	
	We shall also restrict  to actions of definite parity, which is sufficient for equations of motion that preserve parity. This is partly to keep 
	the analysis manageable and partly because we expect all parity-violating $N=4$ CS-like theories to be connected to a parity-preserving 
	theory by a limiting process. However, we must still consider {\it actions} of both positive and negative parity, and allow for all possible intrinsic 
	parity assignments for the auxiliary fields $h$ and $f$. There is no freedom to choose an intrinsic parity for $\omega$; it must have odd parity  because 
	otherwise $R(\omega)$ would have no definite parity. This is also expected from the fact that $\omega$ is the
	{\it dual} spin-connection one-form, i.e.
	\begin{equation}
	\omega^a = \frac{1}{2}\epsilon^{abc} \omega_{bc}	\, ,
	\end{equation}
	where $\omega_{bc}$ is the usual (parity-even) spin-connection one-form. The dreibein one-form $e$ must also have even parity if we insist on even parity for the integral of the
	spacetime volume form $e\cdot e\times e$.
	
	\subsection{Parity-Even Action}
	
	We start our investigation by assuming a parity even action. Irrespective of the choice of intrinsic parity for $h$ and $f$, this requires
	\begin{equation}
	a_{13} = a_{18}  = a_{19} =  a_{20} = 0 \,.
	\end{equation}
	We next observe that the coexistence of some  terms is not permitted by the even parity assumption. For example,  $f\cdot R(\omega)$ and $f \cdot T (e)$ 
	cannot coexist since $T(\omega)$ is parity even while $R(\omega)$ is parity odd; which we allow will
	depend on the choice of intrinsic parity for $f$. This motivates separate consideration of the following possibilities	
	\begin{enumerate}
		\item {$f$ and $h$ are parity odd.}
		\item {$f$ and $h$ are parity even.}
		\item {$f$ is parity odd and $h$ is parity even.}
		\end{enumerate}
		The $4$th case in which $f$ is parity even and $h$ is parity odd is equivalent to case 3.

	\subsubsection{$f$ and $h$ are parity odd}
	
	In this case we require
	\begin{equation}
	a_4 =a_6 =a_8 =a_{10} =a_{11} =a_{12} =a_{14} =a_{15} =a_{17} =0\, . 
	\end{equation}
	As a result, the $e,\omega, h$ and $f$ field equations are given by
	\begin{eqnarray}
	0 &=& a_1 R(\omega) + a_2 e \times e + a_3 f \times f + 
	a_5 h \times h + a_7 f \times h  + a_9 D(\omega) f+  a_{16} D (\omega) h  \,. \nonumber\\
	0 &=&   a_1 T(\omega) +  a_9 e \times f  + a_{16} e \times h  \,,\nonumber\\
	0 &=& a_{16} T(\omega)  + a_7 e \times f + 2 a_5 e \times h  \,,\nonumber\\
	0 &=& a_9 T(\omega)   + 2 a_3 e \times f + a_7 e \times h   \,.
	\end{eqnarray}
	These equations do not allow the simultaneous algebraic elimination of $h$, $f$ and $\omega$. 	
	
	\subsubsection{$f$ and $h$ are parity even}
	
	In this case we require
	\begin{equation}
	a_9 = a_{14} = a_{16} = 0  \,,
	\end{equation}
	which leads to the field equations
	\begin{eqnarray}
	0 &=& a_1 R(\omega) + a_2 e \times e + a_3 f \times f + 2 a_4 e \times f + a_5 h \times h +
	2a_6 e \times h + a_7 f \times h  \,. \nonumber\\
	0 &=&   a_1 T(\omega) + a_8 D(\omega) f   +  a_{15} D(\omega) h   \,,\nonumber\\
	0 &=& 2 a_5 e \times h + a_6 e \times e + a_7 e \times f + 2 a_{11} f \times h + a_{12} f \times f  \nonumber\\
	&& + a_{15} R(\omega)  + 3 a_{17} h \times h \,,\nonumber\\
	0 &=& 2 a_3 e \times f + a_4 e \times e + a_7 e \times h + a_8 R(\omega) + 3 a_{10} f \times f \nonumber\\
	&& + a_{11} h \times h + 2 a_{12} f \times h  \,.
	\end{eqnarray}
	These equations do not allow the simultaneous algebraic elimination of $h$, $f$ and $\omega$. 
	
	\subsubsection{$f$ is parity odd and $h$ is parity even}
	
	In this case we need to set
	\begin{equation}
	a_4 = a_7 = a_8 = a_{10} = a_{11}  = a_{16} = 0 \,,
	\end{equation}
	which reduces the $e,\omega, h, f$ field equations to, respectively, 
	\begin{eqnarray}
	0 &=& a_1 R(\omega) + a_2 e \times e + a_3 f \times f+ a_5 h \times h +
	2a_6 e \times h + a_9 D(\omega) f \,. \nonumber\\
	0 &=&   a_1 T(\omega) +  a_9 e \times f + a_{14} f \times h  + a_{15} D(\omega) h  \,,\nonumber\\
	0 &=& 2 a_5 e \times h + a_6 e \times e  + a_{12} f \times f + a_{14} D(\omega) f  +  a_{15} R(\omega) + 3 a_{17} h \times h \,,\nonumber\\
	0 &=&  2 a_3 e \times f +   a_9 T(\omega)  + 2 a_{12} f \times h + a_{14} D(\omega) h  \,, 
	\end{eqnarray}
	Combining the $\omega$ and $f$ field equations, we find that
	\begin{eqnarray}\label{TorsionDh}
	0 &=&  (a_9 a_{15} - a_1 a_{14}) D(\omega) h +(a_9^2 - 2 a_1 a_3) e \times f + (a_9 a_{14} - 2 a_{12} a_1) f \times h \\
	0 &=& (a_1 a_{14} - a_{9} a_{15}) T(\omega) + (a_9 a_{14} - 2 a_3 a_{15} ) e \times f + (a_{14}^2 - 2 a_{12} a_{15}) f \times h  \nonumber\,.
	\end{eqnarray}
	The first of these equations may be used to express $f$ in terms of $h$ and $D(\omega)h$ but this requires
	a non-zero coefficient for  $e\times f$, and the $f\times h$ term must be absent to avoid an inversion of $(e-h)$ that would lead to an infinite series. 
	The second equation is now the only one involving $T(\omega)$, and must be used to eliminate $\omega$, but this requires
	a non-zero  coefficient for  the $T(\omega)$ term; in addition the  $f\times h$ term must be absent because otherwise $\omega$ will 
	include a torsion tensor that depends on $h$, which now depends on $\omega$, and this would again lead to an infinite series. These
	considerations imply that we should impose 
	\begin{equation}\label{C1}
	a_{14}^2 - 2 a_{12} a_{15} =0 \,, \qquad a_9a_{14} -a_1 a_{12} - a_3 a_{15}  =0\, , 
	\end{equation}
	and 
	\begin{equation}
	a_1 a_{14} - a_{9} a_{15} \neq 0 \,, \qquad a_9^2 -2 a_1 a_3 \neq 0 \, . 
	\end{equation}
	The equations of (\ref{TorsionDh}) then become equivalent to the two equations
	\begin{equation}\label{TorsionDh2}
	T(\Omega) =0\, , \qquad e\times f =   \left(\frac{a_1a_{14}-a_9 a_{15}}{a_9^2-2a_1a_3}\right)D(\Omega) h \, ,  
	\end{equation}
	where
	\begin{equation}
	\Omega = \omega + \alpha f\, , \qquad \alpha = \frac{a_9 a_{14} - 2 a_3 a_{15}}{a_1 a_{14} - a_{9} a_{15}} \, . 
	\end{equation}
	This tells us that $\Omega$ is the usual torsion-free (Lorentz dual) spin connection one-form, and allows us to express
	$f$ in terms of $h$. 
	
	Now we turn our attention to the $e$ and $h$ field equations, which we can rewrite as 
	\begin{eqnarray}\label{e-h-Eqs}
	0 &=& a_1 R(\Omega) + a_2 e \times e + a_5 h \times h + 2a_6\, e\times h +   \gamma D(\Omega) f \, ,\nonumber\\
	0 &=& a_{15} R(\Omega)   + a_6 e \times e + 3 a_{17} h \times h + 2 a_5 e \times h \, ,
	\end{eqnarray}
	where 
	\begin{equation}
	\gamma = \frac{a_{15}(2a_1a_3-a_9^2)}{a_1a_{14}-a_9a_{15}}\, . 
	\end{equation}
	The second of these equations (the $h$-equation) may be solved for $h$, algebraically and finitely (and non-trivially),  provided that 
	\begin{equation}
	a_{17}=0\, , \qquad a_5a_{15}\neq0	 \, . 
	\end{equation}
	Since $a_5\ne0$, there is an $a_5 e\cdot h\times h$ term in the action and a shift of $h$ by a factor times $e$ will allow us to set $a_6=0$ 
	without loss of generality.  Similarly, the freedom to shift $\omega$ by a factor times $f$ in the action, and the fact that $a_{15}\ne0$,  allows 
	us to set $a_{14}=0$ without loss of generality, but it then follows from (\ref{C1}) that $a_3=a_{12}=0$ too. We may therefore set
	\begin{equation}
	a_3=a_6=a_{12}=a_{14}=0\qquad \Rightarrow \quad \alpha=0\, , \quad \gamma=a_9\, . 
	\end{equation}
	We then have
	\begin{equation}\label{hf}
	e\times h= - \frac{a_{15}}{2a_5} R(\omega) \, , \qquad e\times f = -\frac{a_{15}}{a_9} D(\omega)h\, , 
	\end{equation}
	for non-zero $a_5$, $a_9$ and $a_{15}$. The  dreibein equation, which is the first of equations (\ref{e-h-Eqs}), now simplifies to
	\begin{equation}
	0= a_2 e\times e + a_1R(\omega) + 2 a_5 h\times h + 2 a_9 D(\omega)f\, . 
	\end{equation}
	
	The Lagrangian 3-form that yields these equations is
	\begin{equation}
	L= a_1 e\cdot R(\omega) + \frac{1}{3} a_2 e\cdot e\times e + a_9 f\cdot T(\omega) + a_5 e\cdot h\times h + a_{15} h\cdot R(\omega)\, . 
	\end{equation}
	For the choice of coefficients
	\begin{equation}
	a_1=-\sigma, \quad a_2= \frac12 \Lambda_0\, ,\quad a_5=- \frac{1}{2m^2} \, ,\quad  a_9=1\, , \quad a_{15}=-\frac{1}{m^2}\, , 
	\end{equation}
	this Lagrangian three-form coincides with the NMG Lagrangian three-form found in \cite{Hohm:2012vh} except for an interchange of the roles of $h$ and $f$.

	\subsection{Parity-Odd Action}
	
	As in the previous subsection, our starting point  is the Lagrangian three-form (\ref{ewhf}). The assumption of odd parity 
	now  forces us to set
	\begin{equation}
	a_1 = a_2 = a_3 = a_5 = 0 \,.
	\end{equation}
	Furthermore, as in the parity-even case, not all of the remaining terms in (\ref{ewhf}) can coexist. To deal with this we again consider separately the 
	possible parity assignments for $h$ and $f$.
	
	\subsubsection{$f$ and $h$ are parity even}
	
	When both $f$ and $h$ are parity even, we require
	\begin{equation}
	a_4 = a_6 = a_7 = a_8 = a_{10} = a_{11} = a_{12} = a_{15} = a_{17} = 0 \,.
	\end{equation}
	As a result, the $e,\omega,h$ and $f$ field equations are, respectively, 
	\begin{eqnarray}
	0 &=& a_9 D(\omega) f + a_{16} D (\omega) h + 2 a_{20} T(\omega) \,. \nonumber \\
	0 &=&  a_9 e \times f + a_{13} f \times f + a_{14} f \times h \nonumber \\
	&&  + a_{16} e \times h + a_{18} h \times h  + 2 a_{19} R (\omega) + a_{20} e \times e  \,,\nonumber \\
	0 &=&  a_{14} D(\omega) f  + a_{16} T(\omega) + 2 a_{18} D(\omega) h \,,\nonumber\\
	0 &=&   a_9 T(\omega) + 2 a_{13} D(\omega) f + a_{14} D(\omega) h  \,.
	\end{eqnarray}
	These equations cannot be used to determine $f$ and $h$ in terms of $e$ and $\omega$ because that would require at least two equations with
	$e \times f$ and $e \times h$ terms. 
	
	\subsubsection{$f$ and $h$ are parity odd}
	When both $f$ and $h$ are parity odd, we require
	\begin{equation}
	a_7 = a_9 = a_{16} = 0 \,.
	\end{equation}
	As a result, the $e,\omega,h$ and $f$ field equations become, respectively, 
	\begin{eqnarray}
	0 &=&  2 a_4 e \times f +  2a_6 e \times h   + 2 a_{20} T(\omega) \,, \nonumber \\
	0 &=&  a_1 T(\omega) + a_8 D(\omega) f + a_{13} f \times f + a_{14} f \times h +  a_{15} D(\omega) h    \nonumber \\
	&& + a_{18} h \times h + 2 a_{19} R (\omega) + a_{20} e \times e  \,,\nonumber\\
	0 &=&  a_6 e \times e  + 2 a_{11} f \times h + a_{12} f \times f + a_{14} D(\omega) f \nonumber\\
	&& + a_{15} R(\omega) + 3 a_{17} h \times h +  2 a_{18} D(\omega) h \,,\nonumber\\
	0 &=&   a_4 e \times e + a_8 R(\omega)+ 3 a_{10} f \times f + a_{11} h \times h + 2 a_{12} f \times h  \nonumber \\
	&& + 2 a_{13} D(\omega) f + a_{14} D(\omega) h  \,. 
	\end{eqnarray}
	The first of these equations equation can be solved for $\omega$, but the remaining three equations cannot be used to solve for $f$ and $h$ because all
	$e \times f$ and $e \times h$ terms appear in only one of them. 
	
	\subsubsection{$f$ is parity odd and $h$ is parity even}
	
	In this case a parity-odd action requires
	\begin{equation}
	a_6 = a_9 = a_{12} = a_{14} = a_{15} = a_{17} = 0  \,. 
	\end{equation}
	As a result, the $e,\omega,h$ and $f$ field equations become, respectively, 
	\begin{eqnarray}
	0 &=& 2 a_4 e \times f + a_7 f \times h + a_{16} D (\omega) h + 2 a_{20} T(\omega)  \nonumber\\
	0 &=&  2 a_{19} R (\omega) + a_{20} e \times e  + a_8 D(\omega) f  + a_{13} f \times f  + a_{16} e \times h + a_{18} h \times h \nonumber\\
	0 &=&  a_7 e \times f + 2 a_{11} f \times h + a_{16} T(\omega)  + 2 a_{18} D(\omega) h \\
	0 &=&   a_4 e \times e + a_7 e \times h + a_8 R(\omega) + 3 a_{10} f \times f  + a_{11} h \times h + 2 a_{13} D(\omega) f\, . \nonumber
	\end{eqnarray}
	The $e$ and $h$ equations are jointly equivalent to 
	\begin{eqnarray}\label{TorsionDh2}
	0 &=& (4 a_{18} a_{20} - a_{16}^2) T(\omega) + 2( a_7 a_{18} -  a_{11} a_{16}) h \times f + (4 a_4 a_{18} - a_7 a_{16}) e \times f  \\
	0 &=& (a_{16}^2 - 4 a_{18} a_{20}) D(\omega) h + (a_7 a_{16} -4 a_{11} a_{20}) h \times f + 2 (a_4 a_{16} - a_7 a_{20}) e \times f \,. \nonumber \
	\end{eqnarray}
	The first of these equations is now the only one involving $T(\omega)$, and must be used to eliminate $\omega$,  
	and the second equation is now the only one involving $D(\omega)h$, and must be used to eliminate $f$, but this requires 
	\begin{equation}\label{TCons}
	a_7 a_{18} - a_{11} a_{16} = 0 \, , \qquad a_4a_{18} -a_{11}a_{20} =0\, , 
	\end{equation}
	and
	\begin{equation}\label{INEQS}
	4 a_{18} a_{20} - a_{16}^2\neq 0 \,, \qquad a_4a_{16} -a_7a_{20} \neq0\, . 
	\end{equation}
	The equations of (\ref{TorsionDh2}) then become equivalent to the two equations
	\begin{equation}
	T(\Omega) =0\, , \qquad e\times f =   \left(\frac{4a_{18}a_{20} - a_{16}^2}{2(a_4a_{16}-a_7a_{20})}\right)D(\Omega) h \, ,
	\end{equation}
	where
	\begin{equation}
	\Omega = \omega + \beta f\, , \qquad \beta = \frac{4a_4a_{18}    -a_7a_{16}} {4a_{18}a_{20} -a_{16}^2}\, .
	\end{equation}
	This tells us that $\Omega$ is the usual torsion-free (Lorentz dual) spin connection one-form, and it allows us to solve for f in terms of $Dh$. 
	
	Now we turn our attention to the $\omega$ and $f$ equations, which we may rewrite as
	\begin{eqnarray}\label{omf}
	0 &=&  2 a_{19} R (\Omega) + \xi_1 D(\Omega) f  + \xi_2 f \times f + a_{20} e \times e   + a_{16} e \times h + a_{18} h \times h \,,\nonumber\\
	0 &=&   a_8 R(\Omega) + \xi_4  D(\Omega) f  + \xi_3 f \times f  + a_4 e \times e + a_7 e \times h   + a_{11} h \times h \,,
	\end{eqnarray}
	where
	\begin{eqnarray}
	\xi_1 &=&  a_8 - 2 \beta a_{19}    \, , \qquad 
	\xi_2 = a_{13} - \beta a_8 + \beta^2 a_{19}   \,,\nonumber\\
	\xi_4 &=& 2 a_{13} - \beta a_8 \, , \qquad \xi_3 =  3 a_{10} - 2 \beta a_{13} + \frac12 \beta^2 a_{8} \,.
	\end{eqnarray}
	We need to solve some linear combination of these equations for $h$, which means that we need an equation involving $e\times h$, and this equation should not involve $f$ 
	because $f\sim Dh$, and that would lead to a differential equation for $h$.  It should also not have an $h\times h$ term because this will lead to an infinite series solution 
	for $h$. However, the relation $a_7 a_{18} = a_{11} a_{16}$ implies that  the $e\times h$ and $h\times h$ terms cannot be separated by taking 
	linear combinations of the $\omega$ and $f$ equations.  This means that we must  set to zero the coefficient of the $h\times h$ term in at least one of these equations; 
	i.e. either $a_{18}=0$ or $a_{11}=0$.  If we set $a_{18}=0$ then the constraints (\ref{TCons}) and inequalities (\ref{INEQS}) imply that $a_{11}=0$ too. If we instead set $a_{11}=0$ 
	then either $a_{18}=0$ or $a_7=0$,  but the latter option leaves us without an equation containing $e\times h$ but not $h\times h$.  We are therefore forced to choose
	\begin{equation}
	a_{18} = a_{11} = 0 \ \qquad \left(\Rightarrow \ \beta = a_7/a_{16}\right). 
	\end{equation}
	The constraints (\ref{TCons}) are now satisfied, and the first of the inequalities (\ref{INEQS}) reduce to $a_{16}\ne0$. 
	Recalling that $a_{16}$ is the coefficient of $h\cdot T(\omega)$ in the action, we see  that a shift of $\omega$ by a factor times $f$ can be used to set to zero the coefficient $a_7$ 
	of the $e\cdot f\times h \equiv h\cdot f\times e$ term, and a shift of $h$ by a factor times $e$ can be used to set to zero the coefficient $a_{20}$ of the $e\cdot T(\omega)$ term.
	So, without loss of generality, we now set
	\begin{equation}
	a_7=0 \qquad (\Rightarrow \ \beta =0)\, , \qquad a_{20}=0\, . 
	\end{equation}
	
	The $\omega$ and $f$ equations (\ref{omf}) now simplify to 
	\begin{eqnarray}\label{omf-simple}
	0 &=&  2 a_{19} R (\Omega) + a_8 D(\Omega) f  + a_{13} f \times f  + a_{16} e \times h  \,,\nonumber\\
	0 &=&   a_8 R(\Omega) + 2a_{13}  D(\Omega) f  + 3a_{10} f \times f  + a_4 e \times e \,,
	\end{eqnarray}
	where the coefficients are subject to the two inequalities 
	\begin{equation}
	a_{16}\ne0\, ,  \qquad a_4 \neq 0\, . 
	\end{equation}
	We must now combine the $\omega$ and $e$  equations to get an equation involving $e\times h$ for which both the $D(\Omega)f$ and $h\times h$ terms are absent.
	If $a_{13}=0$ this cannot be done, except trivially when $a_8=0$ too,   in which case the $e$ equation reduces to  a quadratic curvature constraint. We therefore 
	exclude $a_{13}=0$, which will allow us to take a combination for which the $D(\Omega)f$ term cancels. Requiring that the $f\times f$ cancels too, but that the $e\times h$ does not,  leads to the 
	additional constraint on coefficients
	\begin{equation}\label{CONSadd}
	3a_8 a_{10} = 2a_{13}^2 \neq 0 \, , 
	\end{equation}
	and the equation
	\begin{equation}
	e\times h= - \frac{1}{2a_{13}a_{16}} \left[ \left(4a_{13}a_{19} - a^2_8\right) R(\Omega)  -2a_4a_8\,  e\times e \right]\, . 
	\end{equation}
	Taking into account the constraint (\ref{CONSadd}), we may write the $e$ equation as 
	\begin{equation}
	0= R(\Omega) + \frac{a_4}{a_8} e\times e + 2\left( \frac{a_{13}}{a_8}\right) D(\Omega) f + 2\left( \frac{a_{13}}{a_8}\right)^2 f\times f\, . 
	\end{equation}
	
	We still have the freedom to normalize the action and the four one-form fields. If we choose to do this by imposing the five conditions
	\begin{equation}
	a_8=1\, , \quad a_{13}=-\frac{1}{2m^2}\, , \quad a_{16}= - m^2\, , \quad a_4= - \frac{\Lambda}{2}\, , \quad a_{19}= - \frac{(m^2+\Lambda)}{2}\, , 
	\end{equation}
	then we recover precisely the EMG action (\ref{EMG-CSL}).

	\section{Discussion}
	
	A peculiarity of  gauge theories in a three-dimensional spacetime is that they may describe massive particles; of any spin $s$
	but  $s>1$ requires a higher-derivative field equation. The simplest $s=2$ example is the parity-violating Topologically Massive Gravity, or TMG, which is a $3$rd-order 
	extension of  3D General Relativity (GR)  propagating  a single massive spin-2 mode.  If we insist on preservation of parity, which implies propagation of a parity 
	doublet of massive spin-2 modes, then the simplest example is ``New Massive Gravity'',  or NMG, which is a 4th-order extension of 3D GR.  Although these massive 3D gravity 
	theories have higher than second-order field equations, they are still second-order in time derivatives, and their linearizations about a Minkowski vacuum are equivalent to 
	the second-order  Fierz-Pauli (FP) equations for a massive spin-$2$ particle (in the NMG case) or the related ``square-root''  FP equations (in the TMG case). 
	
	In contrast to 4D GR, which is  the unique field theory describing interactions of massless spin-$2$ particles at arbitrarily low energy, there can be many 
	inequivalent, but generally covariant, field theories describing interactions of massive spin-2 particles in a 3D spacetime.  For example, there are bi-metric 
	alternatives to NMG, such as ``Zwei-Dreibein Gravity (ZDG).  In fact, NMG  can be viewed as a bi-metric theory but with an auxiliary second `metric'  that can be trivially eliminated. 
	In contrast, the attempt to eliminate the second metric of a bi-metric alternative to NMG leads to a field equation for the remaining metric that involves an infinite series 
	of terms. This last observation suggests that TMG and  NMG could be the  unique 3D gravity theories propagating, respectively, a single massive spin-2 mode or a parity 
	doublet of them, if a restriction is made to field equations given by a {\it finite} number of terms involving a single metric and its derivatives.
	
	As far as we know, TMG and NMG are indeed unique in this sense  if it is additionally assumed that the field equation  arises from variation of an action 
	that also involves only the single metric and its derivatives. This additional assumption appears innocuous but it overlooks the possibility that an equation 
	involving only the metric and its derivatives may be derivable from an action with auxiliary fields that cannot be eliminated from the action, even though
	(by definition of ``auxiliary'') they can be eliminated from the field equations. In this case the field equation, call it  $E_{\mu\nu}=0$,  will be such that
	$\nabla^\mu E_{\mu\nu}\not\equiv 0$, but the inconsistency that this usually entails is absent.  
		
	This new possibility for consistent field equations, dubbed ``third-way'' consistency,  was originally discovered from an exploration of possible modifications of TMG within the
	Chern-Simons-like formulation of massive 3D gauge theories, motivated  by unitarity problems of TMG with an AdS vacuum.  This led to ``Minimal Massive Gravity'', or MMG, 
	which resolves the unitary problems of TMG.  We have shown here that a similar exploration of the possible modifications of  NMG yields a third-way consistent alternative to 
	NMG, but the CS-like action is parity-odd rather than parity even. By analogy with the ``exotic'' parity-odd CS action for 
	3D GR with AdS vacuum, we have called this new massive gravity theory ``Exotic Massive Gravity'', or EMG; the analogy is imperfect but one aspect of it is that, in both cases, 
	the action requires a non-zero cosmological constant term. 
	
	We have also shown that there is a generalization of the CS-like action for EMG to one of no definite parity; a zero cosmological constant is now allowed and linearization 
	about the Minkowski vacuum yields a  $4$th-order equation that is identical to that of linearized  ``General Massive Gravity'', or GMG, so called because it generalizes NMG to 
	allow for arbitrary masses of the two spin-2 modes. As the action for this new parity-violating massive gravity model reduces to the parity-odd EMG action in the limit of 
	equal masses for the two spin-2 modes, we have referred to it as ``Exotic General Massive Gravity'', or EGMG. 
	
	A feature of the CS-like formulation of massive 3D gravity theories is that it greatly simplifies the Hamiltonian formulation, thereby making possible a simple count 
	of the number of local degrees of freedom  independently of any linearized approximation. We have used this method to confirm that the physical phase space of 
	EMG and EGMG  is exactly what one expects from the propagation of  two spin-2 modes in a Minkowski or AdS vacuum. This tells us that there are no additional local 
	degrees of freedom hiding in the non-linearities.

	Although EMG and EGMG both arise naturally within the CS-like formalism as alternatives to NMG and GMG, respectively, we originally found them from a semi-systematic 
	investigation, in the context of 3D gravity,  of third-way  consistent field equations. We have presented a fairly general construction that yields an infinite series of 
	such field equations, of which MMG is the simplest example and EMG/EGMG the next simplest examples; further examples are higher than $4$th-order.	The results
	of this direct construction of  the EMG/EGMG field equations are slightly more general than those of the CS-like action route, in the sense that the parameter space 
	is slightly larger; for example, the EMG field equations remain consistent for zero cosmological constant even though there is no EMG action for this case. 	
	
	Another feature of our general construction of third-way consistent field equations is that it can be  easily modified to generate consistent matter-coupling. 
	The usual procedure, in which the matter stress tensor becomes the source for the metric equation, is inconsistent  if the consistency of the matter-free metric 
	equation is of third-way type; for MMG, for example, the consistent source tensor is quadratic in the matter stress tensor. We have recovered this result very simply
	from our general construction and we have also found the corresponding  matter-coupled extension of EGMG, and of  EMG but in this case the matter coupling 
	violates parity; this is yet another exotic feature of EMG.  
	
	Our  general construction of third-way consistent 3D massive gravity models includes all those with a ``four-flavour'' CS-like action (appropriate to
	the assumption of $4$th-order field equations) on the assumption that  the action is either parity even or parity odd. We have also shown that it includes the 
	EGMG theory for which the action has no definite parity. We have not attempted a systematic analysis of the general parity-violating case, but we suspect that
	such an analysis will not lead to any new theories.   Of course, we are excluding here those 3D massive gravity theories, such as ZDG and its generalizations,  
	that lead to an  equation for the metric that involves an  infinite series of terms in the curvature and its derivatives; one motivation for this exclusion is that the infinite 
	series will likely diverge for  strong fields. 
	
	A unusual property of the EMG and EGMG field equations, given that they involve tensors quadratic in the curvature, is that there is a unique maximally symmetric
	vacuum, which may be de Sitter, Minkowski or anti-de Sitter according to the choice of cosmological parameter $\Lambda$, which is also the cosmological constant.
	In contrast, the cosmological constant for NMG and GMG is a quadratic function of the cosmological parameter.  We have examined the linearization of EMG and 
	EGMG about its AdS vacuum, both at the level of field equations and at the level of the action.  At the level of the field equations, the EGMG parameter space is 
	restricted only by a simple no-tachyon condition, which is satisfied for an AdS radius larger than 
	some critical value that depends on the other parameters. 
	
	Given an action, one can ask whether both modes have positive kinetic energies; this is the no-ghost condition required for 
	unitarity of the perturbative quantum gravity theory.   It is a general feature of parity-preserving 3D field theories, in a Minkowski or an AdS vacuum, 
	that if one mode of a parity doublet is physical then the other mode is also physical if the action is parity-even (i.e. invariant) but a ghost if the action is parity-odd, 
	so EMG cannot be ghost-free.  We have confirmed this and further shown that EGMG 
	also propagates one spin-2 mode as a ghost, so neither EMG nor EGMG is perturbatively unitary.  In addition, and for similar reasons, one of the two central charges 
	of the asymptotic symmetry algebra is negative, implying that the holographically dual 2D conformal field theory, if it exists, is also non-unitary. 
	
	This conclusion is disappointing but we have not considered here the ``critical'' points in parameter space at which any holographical 2D dual would be
	a logarithmic CFT; such CFTs are non-unitary but have applications in condensed matter; see e.g. \cite{Grumiller:2013at}.  We have also not considered
	how the  new ``exotic'' massive gravity models introduced here differ from their ``standard'' NMG/GMG counterparts in a non-relativistic limit; 
	they provide new possible relativistic extensions of the non-relativistic spin-2 theories proposed in the context of bulk properties of fractional quantum 
	Hall states; see \cite{Bergshoeff:2017vjg} for a discussion of this idea with references to the condensed matter literature.  
	
	Finally, since the EMG/EGMG theories admit anti-de Sitter vacua, they also admit Ba\~nados-Teitelboim-Zanelli black holes \cite{Banados:1992wn}, but the
	thermodynamics will be ``exotic'',  as it is for  the exotic Chern-Simons formulation of 3D General Relativity \cite{Townsend:2013ela} and 3D conformal 
	gravity \cite{Afshar:2011qw}. The  new models found here provide a means for exploration of  this topic in the context of massive gravity rather than topological 
	gravity; the EGMG model, in particular, could  be viewed as a  ``massive deformation'' of  3D conformal  gravity.

	\section*{Acknowledgements}  We are grateful to Hamid Afshar,  Stanley Deser and Ozgur Sarioglu
	for helpful correspondence.  The work of PKT is partially supported by the STFC consolidated grant ST/P000681/1.  
	The work of Y.P. is supported by a Newton International Fellowship of Royal Society.

	
	\providecommand{\href}[2]{#2}\begingroup\raggedright\endgroup

\end{document}